\documentclass[11pt,preprint]{aastex}

\newcommand{\e}{\epsilon}
\newcommand{\g}{\gamma}
\newcommand{\ep}{\epsilon^\prime}

\begin{document}

\title{Neutral Beams from Blazar Jets }

\author{Armen M. Atoyan}
\affil{CRM, Universite de Montreal, Montreal H3C 3J7, Canada}
\email{atoyan@crm.umontreal.ca}

\and

\author{Charles D. Dermer}
\affil{E. O. Hulburt Center for Space Research, Code 7653,\\
Naval Research Laboratory, Washington, DC 20375-5352}
\email{dermer@gamma.nrl.navy.mil}

\begin{abstract}
We treat the production of neutrons, photons, and neutrinos through
photomeson interactions of relativistic protons with ambient photons
in the compact inner jets of blazars. Internal synchrotron and
external isotropic radiation due to scattered optical/UV
accretion-disk radiation are considered as target photon fields.  
Protons are assumed to be accelerated to a maximum energy limited by
the size scale and magnetic field of the jet, and by competing energy
losses. We characterize the conditions when the photomeson
interactions of ultrarelativistic protons become effective, and show
that the presence of the external radiation field makes possible
strong energy losses already for protons with energies $E_p\gtrsim
10^{15}$ eV. Without this component, effective energy losses of
protons begin at $E_p\gtrsim 10^{18}$ eV, and would rapidly disappear
with expansion of the blob.

We develop a model describing the production and escape of neutrons
from a comoving spherical blob, which continue to interact with the
ambient external radiation field on the parsec-scale broad line region
(BLR).  Neutrons may carry $\approx 10$\,\% of the overall energy of
the accelerated protons with $E_p\gtrsim 10^{15}$ eV outside the BLR.
Ultra-high energy gamma rays produced by photomeson interaction of
neutrons outside the blob can also escape the BLR. The escaping
neutrons, gamma rays, and neutrinos form a collimated neutral beam
with a characteristic opening angle $\theta \sim 1/\Gamma$, where
$\Gamma$ is the bulk Lorentz factor of the inner jet.  Energy and
momentum is deposited in the extended jet from the decay of neutrons
at distances $l_{d}(E_n)\approx (E_n/10^{17}\;{\rm eV})$ kpc, and
through pair-production attenuation of gamma rays with energies
$E_\gamma\gtrsim 10^{15}$ eV which propagate to $\sim10$-100 kpc
distances. In this scenario, neutral beams of ultra-high energy gamma
rays and neutrons can be the reason for straight extended jets such as
in Pictor A. Fluxes of neutrinos detectable with km-scale neutrino
telescopes are predicted from flat spectrum radio quasars such as 3C
279.
\end{abstract}


\section{Introduction}

Multiwavelength observations of flares from blazars, particularly in
the $\gamma$-ray domain, have convincingly demonstrated that the
compact inner jets of blazars are effective accelerators of particles
to very high energies \citep{har99,wee00}.  Analyses of correlated
X-ray and TeV gamma-ray flares in BL Lac objects lend support to
leptonic models \citep{mk97,cat97,pia98,tav01,kca02}, which imply
efficient acceleration of relativistic electrons in these sources,
probably due to relativistic shocks.  Associated acceleration of
hadrons is expected with at least the same efficiency as that of the
leptons, except perhaps for electron-positron pair jet models where
few hadrons are present. Comparison of the radio lobe and inner jet
powers indicates that jets are composed mainly of electrons and
protons \citep{cf93}, so that a nonthermal hadronic component is
expected.

Acceleration of hadrons in blazar jets could be directly confirmed
with the detection of neutrinos, provided that there are significant
interactions of accelerated hadrons with ambient material or photon
fields. Detectable synchrotron emission at TeV energies could be
 radiated by ultrarelativistic protons and ions, but this requires
 extremely strong magnetic fields $\sim 20$-100 G in BL Lac objects
\citep{aha00,mp01}. All other observable consequences of hadron
acceleration result from the same interactions that produce
neutrinos. These include X-ray and gamma-ray
production from secondary leptons and gamma rays formed in hadronic
interactions and, as we show here, deposition of energy transported by
the escaping neutral radiations far from the inner jet.

Models invoking interactions with ambient matter 
\citep{bb99,ps00,sps02} require mass-loaded jets. If the nuclear
interaction energy-loss time scale $t^\prime_{pp} = (n_p^\prime c
K_{pp} \sigma_{pp})^{-1}$ is to be less than the variability time scale
$t^\prime_{var}$, then high plasma densities $n_p^\prime \gg
10^9/[t_{var}({\rm d})(\delta/10)]$ cm$^{-3}$ are required, where
$n_p^\prime$ is the comoving thermal proton density, $t_{var}({\rm
d})$ is the observed variability time scale in days, $\delta$ is the
Doppler factor, $\sigma_{pp}\cong 30\,$mb is the nuclear
interaction cross section, and $K_{pp}\simeq 0.5$ is the inelasticity.  
(Henceforth primes denote quantities in
the comoving frame.) Such models would, however, be inefficient if the
sources are to be optically thin, as is required for nonthermal X-ray
escape. For Thomson-thin jets ($\tau_{\rm sc} = n_p^\prime \sigma_{\rm
T} R^\prime < 1$), the nuclear interaction time scale
\begin{equation}
t^\prime_{pp} = {\sigma_{\rm T}\over K_{pp} \, 
\sigma_{pp}}\;{R^\prime\over c\tau_{\rm T}} > 40 \;t^\prime_{dyn}\;,
\label{tpp}
\end{equation}
where $R^\prime$ is the comoving blob radius, $t^\prime_{dyn} =
R^\prime/c$ is the dynamical (or light crossing) time scale, and
$\sigma_{\rm T}\cong 665\,$mb is the Thomson cross sections. Thus
protons can only lose $\lesssim 2$-3\% of their energy on the dynamical
time scale. Moreover, as shown in Appendix A, nuclear $pp$ 
interaction models require large masses and kinetic energies.

A second group of hadronic models is based upon photomeson
interactions of relativistic hadrons with ambient photon fields in the
jet. Most of the models of this type take into account collisions of
high-energy protons with the internal synchrotron photons
\citep{mb92,man93,muc02}, while others also take into account external
radiation that originates either directly from the accretion disk
\citep{bp99} or from disk radiation that is scattered by surrounding
clouds to form a quasi-isotropic radiation field \citep{ad01}. BL Lac
objects have weak emission lines, so in these sources the dominant
soft photon field is thought to be the internal synchrotron
emission. Strong optical emission lines from the illumination of
broad-line region (BLR) clouds in flat spectrum radio quasars (FSRQs)
reveal bright accretion-disk and scattered disk radiation in the inner
regions \citep{net90}.

In the case of internal synchrotron radiation, the energy output of
secondary particles formed in photohadronic processes is generally
peaked in the energy range from $\approx 10^{16}$-$10^{18}$ eV in
either low- or high-frequency peaked BL Lac objects \citep{muc02},
which implies that such models can only be efficient if protons are
accelerated to even higher energies. This demand upon proton
acceleration for efficient photomeson production on the {\it internal}
synchrotron photons also holds in FSRQs, which have similar nonthermal
soft radiation spectra as low-frequency peaked BL Lac objects.  As
shown by \citet{ad01}, however, the presence of the isotropic external
radiation field in the vicinity of the jets of FSRQs strongly improves
the photomeson production efficiency and relaxes the very high minimum
proton energies needed for efficient production of secondaries. The
existence of a strong external radiation field is required to explain
\citep{ds93,ds02,sbr94,sik01,dss97,bot00} the luminous 100\,MeV - GeV 
gamma-ray
emission observed with the EGRET instrument on the {\it Compton Gamma
Ray Observatory} \citep{har96}.

In the model of \citet{ad01}, protons are assumed to be accelerated in
an outflowing plasma blob moving with bulk Lorentz factor $\Gamma$
along the symmetry axis of the accretion-disk/jet system. The
relativistic protons are assumed to have an isotropic pitch-angle
distribution in the comoving frame of a plasma blob, within which is
entrained a tangled magnetic field. In our study, we determined the
intensity of the internal radiation fields based on observations of 3C
279 during the flaring state in 1996 \citep{weh98}, and calculated the
high-energy neutrino flux expected under the assumption that the power
to accelerate relativistic protons was equal to the power injected
into nonthermal electrons which explains the observed gamma-ray
emission.  The presence of a quasi-isotropic external radiation field
enhances the neutrino detection rate by an order-of-magnitude or more
over the case where the field is absent, so we predict that FSRQs can
be detected with km-scale neutrino detectors, whereas BL Lac objects
are not as promising for neutrino detection. The model also takes
into account the effects of relativistic neutron production, which can
escape from the blob unless they are converted back to protons due to
decay or further photohadronic collisions inside the blob.

In this paper, we further develop this model. Details of the theory
are presented that describe the evolution of relativistic protons,
taking into account photohadronic energy losses and neutron escape
from the relativistically moving blob. Production of neutrons has been
considered earlier in the stationary cores of AGNs \citep{ew78,sbr89, km89,
brs90, gk90, a92a, a92b}, though not in the context of a jet model.
The neutrons escaping from a relativistic blob form a beam with
opening angle $\theta_n \cong \Gamma^{-1}$. These neutrons are subject
to decay and further photohadronic interactions while passing through
the quasi-isotropic external radiation field on the parsec scale,
which we associate with the broad emission-line region (BLR).
Ultra-high energy gamma rays produced from the decay of secondary
pions outside the blob can escape the BLR because they are no longer
subject to strong $\gamma\gamma$ interactions with the synchrotron
photons inside the blob when they are at a distance of a few $\times
R^\prime$ from the blob.  This results in a neutral beam of ultra-high
energy (UHE) neutrons and gamma rays that can transport the energy to large
distances from the central engine without significant interactions
with the ambient medium until the neutrons decay or gamma rays are
converted to electron-positron pairs through pair attenuation with
diffuse radiation fields. This scenario naturally explains the
appearance of large-scale jets that are colinear with the inner jets 
and appear straight on scales of hundreds of kpc. We also consider whether
the different morphologies of FRI and FRII radio galaxies are a
consequence of the different neutral beam properties formed in BL Lac
objects and FSRQs.

In Section 2, we describe the model, our treatment of photomeson
production and the energy spectra of secondaries, and our method to 
estimate the radiation and magnetic fields in the blob. The results of the 
calculations are presented in Section 3, and a discussion and summary are
given in Section 
4. A comparison of neutrino production through photomeson $p\gamma$ and
nuclear $p p$ interactions in jets is given in Appendix A, and the conditions
when the direct accretion-disk radiation field is less important than
the scattered radiation field are derived in Appendix B.

\section{Model }

Here we consider the production of secondaries formed through
photomeson interactions between relativistic protons and photons of
both the internal synchrotron and the external scattered
accretion-disk radiation fields. The scattered accretion-disk field is
more important than the direct accretion disk radiation field when the
blob is at a distance of hundreds of Schwarzschild radii from the
central engine because, unlike the case for Compton scattering,
 the low-energy threshold for the photomeson
process requires higher energy photons in the rest frame of the
proton. The effects of directionality from photons originating from
the accretion disk make this component less important for photomeson
production than the scattered radiation field. The energy threshold is 
difficult to satisfy for
photons originating from the luminous central regions of the accretion
disk, which enter the plasma blob in a tail-on geometry.  The
accretion disk photons that enter from moderate angles originate 
from cooler portions of the disk at large disk radii, so that
 higher energy protons are required to exceed the photomeson threshold energy
\citep{bp99}. A quantitative estimate of the relative importance of the 
direct and  scattered radiation fields is derived in Appendix B.

For an isotropic spatial distribution of ultrarelativistic neutrons in
the frame of a blob moving with Lorentz factor $\Gamma \equiv
1/\sqrt{1-\beta^2} \gg 1$, the pattern of escaping neutrons in the
stationary frame is beamed with a characteristic opening angle $\theta
\simeq 1/\Gamma$. The beaming of energetic neutrons would become even
more pronounced when one takes into account that the angular
distribution of external radiation is highly beamed in the direction
opposite to the jet in the blob frame.  For isotropic protons, this
will give significant preference to head-on $p\gamma$ collisions of
the protons in the direction of the jet rather than in the opposite
direction.  Therefore, similar to the case of Compton scattering
\citep{der95}, though possibly even more pronounced due to threshold
effects, the external radiation field will cause a significant
directional enhancement of the high-energy neutral beams in the jet
direction already in the comoving frame.  The neglect of this 
 effect does not change, however, the mean energy-loss rate of protons in the 
 blob, which are assumed to be isotropically distributed. It will
 have a noticeable impact only on the angular distribution of secondaries
in near-threshold interactions. We have therefore neglected
this effect in the paper, and a precise treatment of this effect is
deferred to later work. 

In the following subsections, we present our method of calculations of
photomeson and neutron production, proton energy-loss time scales and
the relevant radiation fields, and spectra of secondary leptons,
gamma rays, and neutrinos resulting from the decay of pions. Following
this, we describe how the target photon fields are derived.  We then
present our treatment of proton evolution using a continuity-equation
approach, and develop a theory for the neutron escape and subsequent
neutron and $\gamma$-ray interactions outside the plasma blob.

\subsection{Production of Secondaries in Photomeson Collisions}

Photomeson interactions of relativistic protons with photons of the
ambient radiation field produce secondary electrons and positrons
(referred to henceforth as electrons, unless the distinction is
necessary), gamma rays, and neutrinos primarily through the production
and decay of pions. The relevant decay chains are $\pi^{0} \rightarrow
2\, \gamma$, and $\pi^{\pm} \rightarrow
\mu^{\pm} + \nu_{\mu} \rightarrow e^{\pm} + 2 \, 
\nu_{\mu} + \nu_{e} \, $ 
(we do not distinguish between $\nu_\mu$ and $\bar\nu_\mu$).
It is important that in about half of these inelastic collisions, the 
primary relativistic proton will be converted to a relativistic neutron. 
This is true for the single-pion production channel 
$p + \gamma \rightarrow n + \pi^{+}\;$, which has approximately the same
cross section as the channel $ p + \gamma 
\rightarrow p + \pi^{0}$ without neutron production, as well as for
 multipion production channels. 
  
Energy losses of relativistic protons (and neutrons) are calculated on
the basis of standard expressions (e.g., \cite{bg88}) for the cooling
time of relativistic protons, $t_{p\gamma} = E_p/|dE_p/dt|$, due to
photopion production in $p\gamma$ collisions. If the ambient photons
have spectral density $n_{\rm ph}(\e,\Omega)$ in the direction
$\Omega$, then the cooling time of a proton in the direction
$\Omega_p$ is equal to
\begin{equation}
t^{-1}_{ p \gamma}(\gamma_p,\Omega_p) = 
c \int_0^\infty d\e \oint d\Omega \; (1-\beta_p\cos\psi)\;n(\e,\Omega)
\sigma_{p\gamma}(\e_r) K_{p \gamma}(\e_r)\,,
\label{tpgamma}
\end{equation}
where $\e_r = \e\gamma_p(1-\beta_p\cos\psi)$ is the photon energy in
the rest frame of the proton, $\gamma_{p}$, $\beta_p$c, and $\psi$
are the proton's Lorentz factor, speed, and collision angle with the
photon, respectively, and $ K_{\rm p \gamma}(\epsilon_{\rm r})$ is the
inelasticity of the interaction. The threshold photon energy for pion
production in $p\gamma$ and $n\gamma$ reactions in the rest frame of the 
incident nucleon is $\epsilon_{\rm th} \approx 150 \,\rm MeV$. This means that
protons with, say, 1 PeV energy will effectively interact with soft
X-ray photons with energy $\gtrsim 0.15$ keV.

In the case of an isotropic photon distribution where $n_{\rm
ph}(\e,\Omega) = n(\e)/4\pi$, equation (\ref{tpgamma}) reduces to the form
\begin{equation}
 t_{\rm p \gamma}^{-1}(\gamma_p) =
\int_{\frac{\epsilon_{\rm th}} {2 \gamma_p}}^{\infty}{\rm d}\e\;{
\frac{c \, n_{\rm ph}(\e)}{2 \gamma_p^2 \epsilon^{ 2} }
\int_{\epsilon_{\rm th}}^{2\e \gamma_p}{\rm d} \epsilon_{\rm r}\;{
\sigma(\epsilon_{\rm r}) K_{\rm p \gamma}(\epsilon_{\rm r}) \epsilon_{\rm r} 
  }\, }
\label{tpgiso}
\end{equation}
\citep{ste68,ber90}. The $p\gamma$ collision rate $\nu_{p\gamma}(E_p)$ 
is given by a similar expression as equations (\ref{tpgamma}) or
(\ref{tpgiso}), but with the inelasticity coefficient $K_{p\gamma}$
being dropped from 
the integral. The mean inelasticity $\overline{K}_{p\gamma} (\gamma_p)
= 1/[t_{\rm p \gamma}(\gamma_p)\nu_{p\gamma}(E_p)]$.

A detailed recent study of this photohadronic process is given by
\citet{muc99}. To simplify calculations, we approximate
$\sigma(\epsilon_{\rm r})$ as a sum of 2 step-functions
$\sigma_{1}(\epsilon_{\rm r})$ and $\sigma_{2}(\epsilon_{\rm r})$ for
the single-pion and multi-pion production channels, respectively, 
with $\sigma_{1}=340~ \mu \rm b$ for $ 200 \,\rm MeV \leq
\epsilon_{\rm r}
\leq 500 \, MeV$ and $\sigma_1 = 0$ outside this region, whereas
$ \sigma_2 = 120 ~\mu \rm b$ at $\epsilon_{\rm r} \geq 500 \,\rm MeV$.
The inelasticity in the single pion channel is approximated as $K_{\rm
p\gamma} =K_1\approx 0.2$, whereas $K_{p\gamma}=K_2\approx 0.6$ for
energies above 500 MeV.  Our treatment gives energy-loss pathlengths
of 154, 22.0, and 14.1 Mpc for the photomeson interactions of
ultra-relativistic protons at energies of $10^{20}$, 3$\times
10^{20}$, and $10^{21}$ eV, respectively, with the cosmic microwave
background radiation field. These values are within 2\% of the values
calculated by \citet{sta00}. Even at 7$\times 10^{19}$ eV, where the
interactions occur with the exponential Wien tale of the blackbody
spectrum, our treatment gives an energy-loss pathlength of 650 Mpc,
which is within 10\% of the more accurate values calculated by
\citet{sta00}.  This approach also works well, within a few percent,
for a broad power-law distribution of field photons $u^\prime_{\rm
ph}(\ep )
\propto (\epsilon^{\prime})^{-\alpha}$ for different  spectral 
indices $\alpha$, and readily explains the significant
increase in the mean inelasticity of incident protons (or neutrons)
from $\overline{K}_{p\gamma} \simeq 0.2 $ for steep photon
spectra with $ \alpha \gtrsim 1 $, to
$\overline{K}_{p\gamma} \rightarrow 0.6 $ for hard spectra
with $ \alpha < 1 $, giving results in good agreement with
the treatment of \citet{muc99}.

The spectra of secondary $\pi^{0,\pm}$-decay particles ($\nu,
\; \gamma, \; e$) are calculated in the $\delta$-function approximation 
for the energies of the secondaries. To correctly apply the
$\delta$-function approximation, one has to properly take into account
the different inelasticities of the single-pion and multi-pion
production channels. In the single-pion
production channel, the probability for the conversion of 
a proton to a neutron with the emission of a $\pi^+$-meson is
given by $\xi_{pn} \cong 0.5$. Given that the mean 
energies $E_{\nu}$ and $E_e$ 
of secondary electrons and neutrinos, respectively, are approximately the 
same, and that the $\pi^+$ takes on average $K_1 E_p$ of the initial proton 
energy $E_p$, we have that $E_{\nu}\cong E_e \cong 0.05 E_p$. In the 
single-pion channel, a $\pi^0$-meson is produced with the probability 
$1-\xi_{pn}$,  which results in two gamma-rays
each with mean energy $E_\gamma \cong 0.1 E_p$ after the $\pi^0$ decay.

In the multi-pion channel, we treat the coefficient of the inelasticity 
$K_2 \approx 0.6$ as a result of the production of three leading high-energy 
pions, equally distributed between $\pi^+$, $\pi^-$, and $\pi^0$ mesons, 
which becomes increasingly valid in the limit of large multiplicities. 
These three leading pions are assumed to take most of the energy
lost by the primary proton, so that the remaining pions take away a 
negligible fraction of the initial energy. In this case, the mean energy 
carried by each pion is equal to $0.2 E_p$, and the mean energies carried by 
the secondary electrons, neutrinos, and gamma rays are the same
as in the single-pion channels. 
The production spectra of the secondaries can now be found in the 
$\delta$-function approximation for the differential cross sections for
 parent pion production, given the mean numbers of pions
produced per $p\gamma$ interaction.  

The mean inelasticity $\overline{K}_{p\gamma}= p_1 K_1 + (1-p_1) K_2$
for photomeson interactions of a proton with energy $E_p$ depends on
the interaction probabilities $p_1$ and $p_2 = 1-p_1$ of the proton
with the given photon field via single-pion or multi-pion channels,
respectively. This relation can be inverted to derive the value of
$p_1\equiv p_1(E_p)$, given by
\begin{equation}
p_1 = \frac{K_2 -\overline{K}_{p\gamma}(E_p)}{K_2-K_1} \; . 
\end{equation}
 
Using the numbers of neutral and charged pions produced
in single-pion and multi-pion channels, we can deduce the mean numbers
of high-energy pions produced per $p\gamma$ collision at proton 
energy $E_p$. For $\pi^0$ production we thus have  
$n(\pi^0)=p_1 \, (1-\xi_{pn}) + p_2 = 1 - p_1 \xi_{pn}$, and for the mean
 total number of charged pions $ n(\pi^{+}+ \pi^-) = 2p_2 + p_1\xi_{pn}$. 
Taking further
into account that 2 $\gamma$-rays are produced in $\pi^0$ decay, and
that 3 neutrinos and 1 electron are produced in the decay of every charged
pion, we arrive at the following formulae 
for the production spectra $Q_{x}(E)\equiv {\rm d}N_{x}(E)/{\rm d}t$ 
of gamma rays,  electrons and neutrinos, given by
\begin{equation}
Q_\gamma(E_\gamma) \cong 20 \,(1- \xi_{pn} p_1) \;
\nu_{p\gamma}(10 E_\gamma) \, N_p(10E_\gamma)
\label{Qgamma}
\end{equation}
\begin{equation}
Q_e(E_e) \cong 20 (p_1 \xi_{pn} + 2p_2 ) \,
      \nu_{p\gamma}(20 E_e)N_p(20E_e)
\label{Qe}
\end{equation}
The production spectrum of neutrinos $Q_\nu(E_\nu) = 3Q_e(E_e)$, 
with $E_\nu = E_e$.

\subsection{Photon and Magnetic Fields}

The spectral energy density of the synchrotron radiation field
$u_s^\prime(\ep)$, which is assumed to be isotropic in the comoving
frame, depends on knowledge of the size $R^\prime$ of the emitting
region and the $\nu F_\nu$ spectral fluxes $f_s(\e)$ (erg cm$^{-2}$
s$^{-1}$) measured by an observer during a blazar flare.  We use the
variability time scale to estimate $R^\prime \approx c\delta
t_{var}/(1+z)$ although, strictly speaking, this relation gives 
only an upper limit to the source size.  The comoving synchrotron photon
energy density is then found through the relation
\begin{equation}
 \epsilon^\prime u_s^{\prime}(\epsilon^\prime)\; \cong 
\frac{2 d_{\rm L}^2  f_s(\epsilon)}
{R^{\prime 2} \, c \, \delta^4}\;\cong 
\frac{2 d_{\rm L}^2  (1+z)^2 f_s(\epsilon)}
{ c^3 \, t_{var}^2\delta^6}\; \;,
\label{eq1}
\end{equation}
where $\epsilon = \delta\epsilon^\prime /(1+z)$ relates photon
energies in the observer and comoving frames, and $d_L$ is the
luminosity distance (see, e.g., \citet{tot99}). In our calculations,
we consider an $\Omega_m = 0.3$, $\Omega_\Lambda = 0.7$ cosmology with
a Hubble constant of 65 km s$^{-1}$ Mpc$^{-1}$.

For the spectrum of the external UV radiation field, we
use a \citet{ss73} optically-thick accretion disk radiation field
that is scattered by BLR clouds.  The specific specral energy density 
of the direct disk radiation along the jet axis in this model is given by
     \citep{ds02} 
\begin{equation}
u(\e , \Omega ; h )= {3R_g L_{ad} \over 16\pi^2 c H^3 \tan^3\psi}\;
\delta(\e - \hat\e)\; ,
\label{ss}
\end{equation}
where $L_{ad}$ is the total luminosity of the accretion disk,
$h=H/R_g$ is the height above the accretion disk in units of
gravitational radius $R_g = GM_{bh}/c^2 = 1.5 \times 10^5
M_{bh}/M_\odot \,\rm cm$ for a central black hole with mass $M_{bh}$.
This expression employs a monochromatic approximation for the mean
photon energy $\hat\e = \e_*/r^{3/4}$, where $\e_* \approx 100\,$eV 
depends weakly on $M_{bh}$
and accretion rate \citep{ds93,ds02}, and $r= h\tan\psi $ is the
accretion disk radius (in units of $R_g$) wherefrom originate the
photons $\epsilon$ crossing the jet axis at an angle $\psi$. The
maximum photon energy $\e_{max}$ is given in this approximation by
$\e_*/r_i^{3/4}$, where $r_i$ is the innermost radius of the blackbody
accretion disk, and $r_i \geq 6$ for a Schwarzschild metric.

The spectrum of the isotropic component depends on the integrated disk
field, which has spectral luminosity $L_{ad}(\epsilon ) \propto
\epsilon^{1/3}\exp(-\e/\e_{max})$ from the far IR up to $\e_{max}$.
The quasar 3C 273 displays a pronounced UV bump peaking above $\approx
10$ eV \citep{lic95}.  A thermal emission component is detected in a
low state from the blazar 3C 279 at a temperature of $\sim 20000$ K,
corresponding to $\e_{max} \sim 5$ eV \citep{pia99}. This value,
however, may underestimate the effective maximum temperature due to
uncertainties in the modeling and subtraction of the nonthermal
component. Moreover, in a flaring state, 3C 279 may have a higher
accretion rate and effective accretion-disk temperature.  For
calculations of the spectra of a typical FR II, we take $ \e_{max} =
20$ eV. For $\epsilon_\ast \approx 100 \,\rm eV$ this implies an 
effective disk innermost radius $r_i\simeq 8$.  

The energy density of the quasi-isotropic component of the surrounding
radiation field can be approximated by the expression
\begin{equation}
u_{ext}(\e ) \cong {L_{ad}(\e ) \tau_{\rm T}\over 2\pi R_{BLR}^2 c}\;,
\label{uext}
\end{equation} 
where $R_{BLR}$ is the effective radius of the BLR, and $\tau_{\rm T}$ is
the Thomson depth of the BLR.\footnote{Note that the factor $2\pi$ in
the denominator of equation (\ref{uext}), instead of the commonly used
$4\pi$, corresponds to the fact that we are considering the photon
density not outside, but inside the emission region \citep{aa96}.} 
The energy density of the external radiation field in the jet frame is
$u^\prime_{ext}\cong \Gamma^2 u_{ext}$, where $u_{ext} = \int_0^\infty
d\e \, u_{ext}(\e )$.

The quantities $L_{ad}$, $\tau_{\rm T}$, and $R_{BLR}$ which define
the energy density of isotropic external field component in equation
(\ref{uext}) are not well-known for specific sources.  The
quantity $u^\prime_{ext}$ can also be estimated from the ratio of the
$\nu F_\nu$ peak flux $f_{EC}^{pk}$ of the $\gamma$-ray Compton
component to the peak flux $f_{s}^{pk}$ observed in the lower
frequency nonthermal synchrotron component.  Assuming that the
observed 100 MeV - GeV fluxes as observed with EGRET from FSRQs in the
flaring state are primarily due to Compton-scattered radiation from
external photon fields, then
\begin{equation}
u^\prime_{ext} \cong gu^\prime_B {f_{EC}^{pk}\over f_{s}^{pk}}\; , 
\label{uprimeext}
\end{equation}
where $u^\prime_B = B^{\prime 2}/8\pi$ is the comoving magnetic-field
energy density, and $g \gtrsim 1$ is a parameter that corrects for the
Klein-Nishina effect in the observed 100 MeV - GeV
fluxes.\footnote{This estimate does not take into account the
different beaming factors for the two processes \citep{dss97}, but is
valid when $\Gamma \cong \delta$.}  For definiteness, note that the
Klein-Nishina transition occurs when the parameter $b =
4\gamma\epsilon^\prime_0/(m_ec^2)$ exceeds unity \citep{bg70}. This
implies that 20 eV UV target photons will be scattered in the
Klein-Nishina regime in order to produce gamma rays with energies
exceeding $\approx 800/(1+z)$ MeV (independent of $\Gamma$). As shown
by our numerical calculations, this requires that $g\sim 3$ for the
parameters of 3C 279 that we consider.

The magnetic field $B^\prime$ in the blob can be determined by
introducing the equipartition parameter $\eta =
u^\prime_{e}(1+k_{pe})/u^\prime_B$ for the ratio of relativistic
electron to magnetic-field energy densities in the jet, with the
factor $k_{pe}=u^\prime_p/u^\prime_e$ correcting for the contribution
of nonthermal hadrons to the particle energy density.  We assume
$k_{pe}$ = 1 in the estimate of $B^\prime$.  The energy density in
nonthermal electrons is inferred from the measured synchrotron flux
density $F_s(\nu ) \propto
\nu^{-\alpha}$ in the range $\nu_0\leq \nu\leq \nu_1$, with $\alpha\cong 0.5$. 
This gives the magnetic field
\begin{equation}
B^\prime({\rm Gauss})\cong 130 \;{d_{28}^{4/7}f_{-10}^{2/7}[(1+k_{pe})
\ln(\nu_0/\nu_1)]^{2/7}(1+z)^{5/7}\over \eta^{2/7}[t_{\rm var}({\rm
d})]^{6/7}\delta^{13/7}\nu_{13}^{1/7} }\;,
\label{Beq}
\end{equation}
and the equipartition magnetic field corresponds to the parameter
$\eta = 1$. Here $\nu_{13}\equiv \nu/10^{13}$ Hz is the frequency
where the flux $f_s(\e )\equiv 10^{-10} f_{-10} {\rm~ergs~cm}^{-2}{\rm
~s}^{-1}$ is measured.

An alternative estimate of $B^\prime$ can be derived from the ratio of
the synchrotron self-Compton (SSC) $\gamma$-ray and synchrotron peak
fluxes $f^{pk}_{SSC}/f^{pk}_s \cong u^\prime_{s}/u^\prime_B$
\citep{sik97,tav98}.  Thus $u_B^\prime = A u^\prime_s =
Af^{pk}_s(2d_L^2/R^{\prime 2} c \delta^4)$, where $A \equiv
(f^{pk}_s/f^{pk}_{SSC})$.  This gives
\begin{equation}
B^\prime({\rm Gauss}) = {4 \sqrt{\pi}(1+z)d_L\over c^{3/2}\delta^3
t_{var}}\;\sqrt{A f_{s}^{pk }} \cong 1.6\;{d_{28}(1+z)\sqrt{A
f_{-10}}\over (\delta/10)^3 \;t_{\rm var}({\rm d})}\;.
\label{B2}
\end{equation}
This equation assumes that Klein-Nishina effects for the SSC $\gamma$
rays are unimportant, because the Compton process occurs on softer
(radio-optical) photons as compared to the UV photons for external
Compton scattering.  In both equations (\ref{Beq}) and (\ref{B2}), the
strongest dependence of $B^\prime$ is through the unknown Doppler
factor. By equating these two equations, we can derive an estimate for
$\delta$, given by
\begin{equation}
\delta \cong 8.9\; {d_{28}^{3/8}\,(1+z)^{1/4}\, f_{-10}^{3/16} \,
 A^{7/16} \, \eta^{1/4}\, \nu_{13}^{1/8}\over
[(1+k_{pe})\ln(\nu_1/\nu_0)]^{1/4}\; [t_{var}({\rm d})]^{1/8}}\;.
\label{deltaequal}
\end{equation}

\subsection{Energy Distribution of Protons and of Escaping Neutrons}

Relativistic neutrons, being essentially unaffected by magnetic
fields, can escape from a blob if their Lorentz factor $\gamma_n$ is
sufficiently large so that they do not decay into protons on the decay
timescale $\tau_{\rm d}(\gamma_n) =\tau_0 \, \gamma_n $, where $\tau_0
\simeq 910 \,\rm s$ is the decay time of a neutron at rest. Moreover, 
the neutrons must not be converted back to protons in further 
photomeson interactions with ambient photons inside the blob during
the crossing time $t_{\rm cross}\simeq R^\prime/c$ needed, on average,
for a neutron to leave the blob.  This leakage must be taken
into account in the kinetic equation describing the evolution of the
proton energy distribution $N_{\rm p}(E,t)$ in the blob at time $t$
(allowing for the possibility of a non-stationary source). We develop a
formalism for the evolution of the proton distribution that properly
treats the possible escape of protons while in the neutron state.

The distribution $N_{\rm p}(E,t)$ can be calculated in the framework
of the approximation of continuous energy losses. Even though
inelasticities up to $K_{p\gamma}\cong 0.5$ can be reached in
photomeson interactions, the continuous energy-loss approximation is
reasonable as follows from the analytic solution for the kinetic
equation of relativistic particles when both continuous and
catastrophic energy losses are taken into account \citep{a92b}.  The
kinetic equation for $N_p(E,t)$ is then given by
\begin{equation}
\frac{\partial N_{\rm p}}{\partial t} = 
\frac{\partial}{\partial E}
 \left( P_{ p\gamma} \, N_{\rm p} \right) - 
\frac{N_{\rm p}}{\tau_{\rm esc}} + Q_{\rm p} \, ,
\label{N_p}
\end{equation}
where $Q_{\rm p}\equiv Q_{\rm p}(E, t)$ is the particle injection rate,
and   $P_{p\gamma}(E) = -(dE/dt)_{p\gamma}$
is the proton energy loss rate due to $p\gamma$ interactions.  
Note that $P_{p \gamma}$ includes
{\it both} the channels with and without conversion of the incident proton
to the neutron, namely $p+\gamma \rightarrow p +\pi^0 +X$ and $p+\gamma 
\rightarrow n +\pi^{+} +X$. This can be done because in 
photomeson interactions the energy loss rate of a
{\it neutron} that is produced in the latter interaction chain 
is practically the same as of the proton in the first chain. 
Therefore once the nucleon
returns to the proton pool inside the blob, it does not `remember' 
that during some time $\Delta t < t_{\rm cross} $ it was in the `neutron'
state. In other words, the energy loss of protons remaining in the blob
 would depend
only on the overall time they spend in the blob after their injection.

The parameter $\tau_{\rm esc}\equiv \tau_{\rm esc}(E)$ corresponds to
the characteristic escape time of protons from the blob. This should
generally take into account the escape of protons both due to direct
escape from the blob while in the neutron state, as well as due to
diffusive escape. In this section, we neglect diffuse escape effects,
although this process is taken into account to determine the spectra
of protons at the highest energies in the numerical calculations. The
solution to equation (\ref{N_p}) is 
\begin{equation} N_{\rm p}(E,t) =
\frac{1}{P_{p\gamma}(E)} \int_{-\infty}^{t} P(\varepsilon_{t_1})
Q(\varepsilon_{t_1},t_1) \exp
\left( -\int_{t_1}^{t}\frac{{\rm d} x}
{\tau_{\rm esc}(\varepsilon_x)}\right)
 d t_1\, .
\label{N_psolt}
\end{equation}
The  parameter $\varepsilon_{t_1} \equiv \varepsilon(E,t,t_1)$ 
corresponds to the initial energy of a particle at an 
earlier time $t_1$ which cools down to a given energy $E$ by the
time $t$; thus $\varepsilon(E,t,t_1=t) =E$. It is determined from 
the equation
\begin{equation}
t-t_1 =\int_{E}^{\varepsilon_{t_1}}{\frac{{\rm d}E}
{P_{p\gamma}(E)}} \, .
\label{tt1}
\end{equation}
Note that the running time $t_1$ can be formally both smaller and 
larger than the time $t$ at which the particle energy $E$ is given, i.e.,
$\varepsilon$ is simply a trajectory of such particle in the energy space.

The characteristic escape time of protons can be evaluated if we
recall that $\Delta N_{\rm esc} = (N_{\rm p}/\tau_{\rm esc})\times
\Delta t$ is the number of particles escaping the source during the
time interval $\Delta t$. It can be written also as $\Delta N_{\rm
esc}/\Delta t = \xi_{\rm esc} \dot{N}_{\rm n} = \xi_{\rm esc} \xi_{p
n} \nu_{p\gamma}(E)\, N_{\rm p}(E,t)$, where $ \xi_{p n} \equiv \xi$
($\cong 0.5$) is the probability of conversion of a proton into the
neutron per one $p \gamma$ collision, $\nu_{p\gamma}$ is the frequency
of such collisions, and $ \xi_{\rm esc}$ is the probability that the
produced neutron would escape the blob prior to its conversion back to
the proton state. Thus
\begin{equation}
\frac{1}{\tau_{\rm esc}(E)} =  \xi_{\rm esc}(E) \, 
  \xi_{p n} \, \nu_{p\gamma}(E) \; .
\label{tauesc}
\end{equation}

In order to calculate $\xi_{\rm esc}$, consider the solution to the equation 
\begin{equation}
\frac{\partial N_{\rm n}}{\partial t} = 
\frac{\partial}{\partial E}
 \left( \xi_{n n} P_{ n\gamma} \, N_{\rm n} \right) - 
(\xi_{np} \nu_{n\gamma} +1/\tau_{\rm d} )\, N_{\rm n} 
+ Q_{\rm n} \, 
\label{partialN}
\end{equation}
for the energy distribution of neutrons $N_{\rm n}(E,t)$ in the blob,
assuming instantaneous injection at time $t_0=0$, which corresponds to
the source function $Q_{\rm n}(E,t) = N_0(E) \delta(t-t_0)$.  In this
equation, $P_{n\gamma} $ and $\nu_{n\gamma}$ represent the overall
photomeson energy loss and collision rates of the neutrons,
respectively, which are approximately the same as the respective rates
for the proton photomeson interactions. Unlike in eq.\ (\ref{N_p}) for
the protons, however, we now have to take into account that after each
photomeson collision, a neutron either remains neutron with
probability $\xi_{n n}\; (\cong 0.5)$ while losing energy, or is
converted to a proton and returns to the proton pool with probability
$\xi_{n p}= 1-\xi_{nn}$.  The neutron escape term also takes into
account the decay of neutrons $n\rightarrow p + e^{+} +\nu$ on
timescale $ \tau_{\rm d} = \tau_{\rm d}(E)$, with $E_n \cong E_p$.

The solution to equation (\ref{partialN}) is given in general by
equation (\ref{N_psolt}). For the instantaneous injection function
$Q_{\rm n}(E,t) = N_{0}(E)\delta(t)$, equation (\ref{N_psolt}) reduces to
\begin{equation}
N_{\rm n}(E_1,t_1) =  
\frac{P_{n\gamma}(E) \, N_0(E)}{P_{p\gamma}(E_1)} 
 \exp \left[ -\int_{0}^{t_1}
\left(\frac{1}{\tau_{\rm d}(E_x)}+\xi_{np}\nu_{n\gamma} \right)
{\rm d}x \right] \; .
\end{equation} 
Here we use $E_1$ for the running energy at $t_1$, while
reserving $E$ for the initial neutron energy. Note that in the continuous
energy loss-approximation, $E$ should be the same as the energy of 
the parent proton. 

Because the neutrons escaping the source are the ones which
survive during the time $t_1\geq t_{\rm cross} \simeq R^\prime/c$ after 
injection, the escape probability $\xi_{\rm esc}$ is formally the ratio 
$N_{n}(E_1, t_1) \Delta E_1 / N_0(E) \Delta E$ at time
$t_1 = t_{\rm cross}$. Taking also into account that,
using equation (\ref{tt1}), the ratio of the energy intervals   
$\Delta E_1 / \Delta E = P(E_1)/P(E)$, we find that
\begin{equation}
\xi_{\rm esc}=  \exp \left[ -\int_{0}^{t_{\rm cros}}
\left(\frac{1}{\tau_{\rm d}(E_x)}+\xi_{np}\nu_{n\gamma}(x) \right)
{\rm d}x \right] \; .
\end{equation}
This defines the effective time of proton escape $\tau_{\rm esc}(E)$
via equation (\ref{tauesc}).  The evolution of neutrons outside the
blob is given by equation (\ref{partialN}) with the source function
$Q_n \equiv Q_n(E,t)$ equal to the rate of escaping protons from the blob in
equation (\ref{N_p}), i.e., 
$Q_{n}(E,t) = N_{p}(E_p,t)/[(1-\overline{K}_{p\gamma}) \tau_{esc}(E_p)]$,  
where 
$E_p= E/(1-\overline{K}_{p\gamma})$ is the mean energy of the protons
producing neutrons with energy $E$.

\section{Results of Calculations}

We calculate the spectra of protons, neutrons, and secondaries from
photomeson interactions, taking inner jet parameters derived from
quasi-simultaneous observations \citep{weh98} of the 1996 flare from
3C 279 as typical for FSRQs.  For calculations of the photomeson
production effects in low luminosity blazars, we use parameters of the
blob derived from flaring states of the BL Lac object Mrk 501
\citep{cat97,pia98}.

\subsection{Photopion Production in FSRQs}

The redshift of 3C 279 is $z = 0.538$, which
implies a luminosity distance $d_{\rm L}\cong 1.05\times10^{28}\,\rm
cm$ for an $\Omega_m = 0.3$, $\Omega_\Lambda = 0.7$ cosmology with a
Hubble constant of 65 km s$^{-1}$ Mpc$^{-1}$.
We use parameters derived for the flaring state of 3C 279 during
1996 February 4-6 \citep{weh98}. Note that parameters appropriate to 
the extended 3 week quiescent state of the same observing period
 were used earelier in \citet{ad01}.
The measured variability time scale $t_{var}\cong 1$ day for the 1996 flare
of 3C 279 implies a characteristic source size 
\begin{equation}
R^\prime \cong ct_{var}\delta/(1+z)\cong 1.7\times 10^{16}
t_{var}({\rm d})(\delta/ 10)\;{\rm ~ cm}\;.
\label{Rprime}
\end{equation}

For calculations of $n^\prime_s(\ep )=u^\prime_s(\ep )/\ep$ in
equation (\ref{eq1}), we approximate the flux density
$F_s(\e)=f_s(\e)/\e\propto \e^{-\alpha}$ observed during the flare of 1996
\citep{weh98} in the form of a continuous broken power-law function,
with indices $\alpha_{1}\cong 0.5$ at frequencies $\nu_0 \cong
10^{11}{\rm ~Hz}< \nu=\e/h \le \nu_1 =10^{13}{\rm ~Hz}$,
$\alpha_{2}\cong 1.45$ at $\nu_1 \leq \nu \leq \nu_2 =10^{16}\,\rm Hz$, 
and $\alpha_{3}\cong 0.6$ at $\nu \geq \nu_2$.  The $\nu F_\nu$
synchrotron radiation flux reaches its maximum value
$f_s^{pk}=f_s(\epsilon_{pk}) \cong 1.7\times 10^{-10} \,\rm erg\,
cm^{-2}\, s^{-1}$ at $\e = \e_{pk} = h\nu_1$. The flaring flux of the
Compton $\gamma$-ray component peaks at $\sim 500 \,\rm ~MeV$, and is
$\sim 15$ larger than $f_s^{pk}$ during the flare \citep{weh98}.

Spectral models for 3C 279 show that a complete spectral fit requires
that the ratio of synchrotron to SSC peak fluxes $A \simeq 1$
\citep{bot99,har01}.  This implies from equation (\ref{deltaequal})
that $\delta \simeq 6$, which is just above the lower limit $\delta
\geq 5$ required from the condition that the gamma rays in the jet be
transparent to $\gamma\gamma$ pair attenuation \citep{ad01}. In the
calculations, we treat a range of Doppler factors $\delta = 6,$ 10 and
15, noting that there can be uncertainties in our knowledge of $A$, as well
as departures from equipartition.  Note, however, that the range of the
Doppler factors cannot be very large due to the weak dependence of
$\delta$ on the uncertain parameters. The magnetic field is then
derived from equation (\ref{Beq}).

During the
1996 February 4-6 flare, a mean apparent isotropic `$4\pi$' $\gamma$-ray
luminosity of $10^{49}$ ergs s$^{-1}$ was measured with
EGRET. Assuming that the power
injected in relativistic protons is equal to the power observed in
$\gamma$ rays, we take for the comoving proton luminosity $L_p^\prime
= 10^{49}/\delta^4$ ergs s$^{-1}$.  The injection of protons takes
place during the time $t_{inj} = 2$ days in the observer frame,
corresponding to the time $\delta t_{inj}/(1+z)$ in the comoving blob
frame. For the spectrum of protons, we take $Q_p(E_p)
\propto E_p^{-2}\exp(-E_p/E_{max})$, where $E_{max}$ is defined by 
the condition that the proton gyroradius 
does not exceed $R^\prime$ and that the proton cooling rate
$t_{p\gamma}^{-1}(E_p)$ does not exceed the maximum acceleration rate
given by $eBc/E_p$.  This acceleration rate 
is consistent with the maximum energy of particles accelerated 
by relativistic shocks \citep{ach01}.
The BLR radius $R_{BLR}$ is deduced from equation
(\ref{uext}) with the assumption $\tau_{\rm T} = 0.1$. Equation
(\ref{uprimeext}) is used to estimate the external radiation density
$u^\prime_{ext}(\ep )$ in the blob frame, noting that $\ep
u^\prime_{ext}(\ep) \cong \Gamma^2 \e u_{ext}(\e )$ (e.g., Dermer \&
Schlickeiser 2002).  This approach results in a dependence of the BLR
radius on $\delta $ (which we assume to be equal to $\Gamma$),   $R_{BLR}
= 0.11$ pc and 0.47 pc for $\delta = 6$ and $\delta = 10$,
respectively.  For the range of Doppler factors used in our
calculations, the blob remains within the BLR during the entire time
of injection of relativistic protons.

Figure~1 shows the spectra of protons and escaping neutrons and gamma
rays for $\delta = 6$ and $\delta = 10$ in Figures 1a and 1b,
respectively. The dotted curves show the overall spectra of protons
injected into the blob, and the thin solid curves show the
spectra of protons which remain in the blob by the time that the blob
reaches the edge of the BLR at stationary-frame time $t =
R_{BLR}/c$. The thick solid curves show the overall spectra of neutrons
which escape from the relativistic blob during the time of its
propagation through the BLR. The feature in the neutron spectrum
between $10^{15}$ and $10^{16}$ eV in Figure~1a and, to a lesser
extent, in Figure 1b, is due to the neutron energy losses on the
external radiation field which are only significant above $\sim
10^{15}$ eV.
The magnetic fields for the $\delta = 6$ and $\delta = 10$ cases in 
Figures 1a and 1b are equal to 18.1\,G and 7\,G, respectively.
The corresponding characteristic maximum energies of the accelerated protons 
in these two cases are equal to $2.5\times 10^{18} \,\rm eV$ and
$6.5\times 10^{18}\,\rm  eV$ in the blob frame, which are translated 
to energies by a factor of $\Gamma \simeq \delta$ times higher in the 
stationary frame.  

Neutrons escaping from the blob moving with Lorentz factor $\Gamma$
will form a highly collimated beam with opening angle $\theta \sim
1/\Gamma$. These neutrons continue to interact with the external
radiation field through photopion production, and are also lost
through decay.  Both effects modify the neutron spectra until the
neutrons reach the edge of the BLR with the external UV field.  
The spectra of surviving neutrons
reaching this edge are shown by the
dashed curves. Photomeson interactions of neutrons outside the blob
result in the production of a beam of ultra-high energy neutrinos,
electrons, and $\gamma$-rays with the same opening angle $\sim 1/\Gamma$.  
The expected number of neutrinos to be detected with ground-based neutrino
telescopes is calculated below. The electrons may be deflected by the
magnetic field in the BLR to produce synchrotron or Compton-scattered
radiation, but we do not consider this emission further here because
the fluxes will depend significantly on unknown properties of the
medium outside the jet, such as the magnetic field strength and
geometry.

The spectra of $\pi^0$-decay gamma rays produced by neutrons outside
the blob which escape from the BLR without undergoing subsequent
$\gamma\gamma$ interactions are shown by the dot-dashed curves in
Figure 1. These gamma rays have very high energies, ranging from
$10^{14}$ to $10^{18}$ eV. The thin curves in Figure 2 show the
$\gamma\gamma$ pair-production opacity due to interactions of gamma
rays with photons of the BLR, calculated for a gamma ray traversing
the radius of the BLR. Also shown by the heavy curves are the
$\gamma\gamma$ opacities within the blob. This figure demonstrates
that gamma rays produced within the blob are subject to very strong
attenuation, and therefore have no chance to escape. The attenuation
is also very strong for photons produced outside the blob in the BLR
unless these photons have very high energies ($\gtrsim 10^{16}$ eV) or
are produced close to the edge of the BLR. The peak
feature in the escaping
$\gamma$-ray spectrum between $10^{14}$ and $10^{15}$ eV in Figure~1a
is due to the threshold character of photomeson interactions of
neutrons with the external radiation field, which continue to produce
gamma rays until they reach the edge of the BLR.

We calculate the fraction of energy initially injected in the form of
protons above $10^{15}$ eV, $W_p(>10^{15}{\rm eV})$, which is taken
away from the blob in the form of neutrons. For the cases of $\delta =
6$ and 10, as shown in Figure~1, these fractions are $14.8\%$ and $25.9\%$,
respectively.  Neutrons with energies greater than $10^{17}$ eV that
escape from the the BLR will decay on length scales
\begin{equation}
l_d \approx 1\;(E_n/10^{17})\;{\rm kpc}
\label{ld}
\end{equation}
The fraction of injected relativistic proton energy $W_p(>10^{15}{\rm eV})$ 
carried by these neutrons, which we denote by $W_n(>10^{17}{\rm
eV})$, is equal to $0.00059\,\%$ and $3\%$ for $\delta = 6$ and 10,
respectively, for the parameters used in Figure~1. The fraction of
ultrarelativistic energy that escapes from the BLR in the form of
gamma rays above $10^{15}$ eV is equal to $ 0.23\%$ and 5.2\% for
$\delta = 6$ and 10, respectively. The fact that a smaller fraction of
initial energy is taken away by neutrons for $\delta = 6$ than for
$\delta = 10$ is explained by the stronger external radiation field in
the former case. This is a consequence of the strong dependence of the
inferred external radiation field energy density $u_{ext}^\prime
\propto \delta^{-26/7}$, as follows from equations (\ref{uprimeext})
and (\ref{Beq}), which strongly attenuates the beam of neutrons
escaping the blob close to the black hole. Our calculations show that
the fraction of injected energy in ultrarelativistic protons,
$W_p(>10^{15}{\rm eV})$, that escape the BLR in the form of neutrons
and gamma rays can reach $\sim 10$\% for different model parameters,
when attenuation of the neutrons escaping the BLR is not very strong.
In particular, the yield in neutrons leaving the BLR is  
always high if the proton acceleration in the blob occurs not during a short
flaring period, as in Figure 1, but during the time 
$t \sim t_{BLR}= R_{BLR}/c$ 
when the blob is traveling through the isotropic external radiation field.
On the other hand, a higher rate of degradation of neutrons results in
higher fluxes of neutrinos produced both inside and outside the
blob. Figure 3 shows the fluences of neutrinos integrated over several
days in the observer frame determined by the time for the blob to pass
through the BLR, for the same parameters as used in Figure 1. The solid
and dashed curves show the fluences calculated for $\delta = 6$ and
10, respectively. The thick and thin curves represent the fluences of
neutrinos produced by photopion interactions inside and outside the
blob, respectively. For the spectral fluences shown in Figure 3, the
total number $N_\nu$ of neutrinos that could be detected by a 1 km$^3$
detector such as IceCube, calculated using neutrino detection efficiencies
 given by \citet{ghs95}, are 0.29 and 0.078, for $\delta = 6$, and 0.13 and
0.076, for $\delta = 10$, where the pair of numbers refer to neutrinos
formed inside and outside the blob, respectively.

The dot-dashed and the triple-dot-dashed curves represent the neutrino
fluences that would be expected for the same time scales and source
parameters, but neglecting interactions with the external radiation
field. The numbers of neutrinos that are predicted to be detected by a
1 km$^3$ detector in this case are more than an order-of-magnitude
lower, and are 0.030 and 0.0047 for $\delta = 6$ and $10$,
respectively. As previously noted \citep{ad01}, the presence of
external radiation fields is crucial for prospects to detect neutrinos
from blazars with high-energy neutrino telescopes.
 
In order to understand how the production of neutrons will change when
the inner jet expands, we show in Figures 4a and 4b the spectra of
protons and escaping neutrons and gamma rays, as in Figure 1, but calculated for
different sizes of the blob corresponding to $t_{var} = 0.3$ and 3
days, with $\delta = 10$. Here we have fixed the external radiation
field density and size of the BLR ($R_{BRL} = 0.41$ pc) to the values
derived in Figure 1b for $\delta = 10$.  
Moreover, we assume that the acceleration rate of protons is
such that the same amount of total energy as in Figure 1
is injected in the protons continuously during the entire period that 
the jet travels through the BLR.
The fractional energy taken
away by neutrons from the compact blob and from the BLR is $\approx
30\%$ and $\approx 7.5\%$, respectively, for $t_{var} = 0.3$ d.  In
the case of the inflated blob with $t_{var} = 3$ d, these fractions
drop down correspondingly to $12$\% and $2.4$\%.

In Figure 5 we show the radiation fluxes produced in and escaping from
the blob following the electromagnetic cascade initiated by energetic
pion-decay electrons and gamma rays for the
case $\delta = 10$. Figure 5a shows the cascade radiation spectra at
the maximum, which is reached at time $t=2\,\rm d$ when the
injection of newly accelerated protons in the blob stops. Figure 5b shows the
radiation fluxes at the time when the blob is leaving the external
radiation field region. The thick and thin curves correspond to
synchrotron and Compton-scattered radiation, respectively. The
radiation of the first generation of electrons, which includes both
the electrons from $\pi^\pm$ decays and the electrons produced by
absorption of $\pi^0$-decay gamma rays in the blob, are shown by the
solid curves. The synchrotron radiation strongly dominates the
Compton-scattered emission in the first generation because the
electrons of the first generation are ultra-relativistic,
such that Compton scattering is in the extreme Klein-Nishina regime.  
The second, third, and fourth generations of the
cascade radiation are shown by the dashed, dot-dashed, and triple-dot
dashed curves, respectively.  In the calculations of the 
escaping cascade radiation spectrum, we have not considered 
 the emission from electrons which are produced outside the blob.
 These electrons become isotropized and produce much weaker observed
 fluxes than the Doppler-boosted
 emission produced in the blob. 
Most of the $\tau_{\gamma\gamma}$
opacity is due to the external radiation field above several GeV 
(see Figure 2).
Thus the spectra of different generations of cascade radiation
escaping the BLR are cut off at nearly the same energy as long as
the blob is in the BLR. At the same time, the cascade radiation 
made when the blob is outside the BLR can extend to higher energies.

The heavy dots show the total
spectrum. 

As can be seen in Figure 5, emission near 1 GeV can escape from the
blob and BLR.  These figures show that in both cases the main
contribution to the resulting gamma-ray flux is due to synchrotron
radiation of the first generation of the cascade electrons born mostly
with energies above $10^{14}\,\rm eV$. 

Noticeable gamma-ray fluxes may also be produced by secondaries
from $p\gamma$ interactions in the blob at later stages
 when the blob is still within the BLR, as seen 
in Figure 5b. One may also expect that $\sim 10\,$GeV gamma-ray
photons produced at these later stages can be due to the escaping 
neutrons interacting with the UV photons outside of the BLR.
 
\subsection{Photopion Production in BL Lac Objects}

BL Lac objects have generally lower luminosity than FSRQs and have
weak or absent line emission, suggesting that the column density of
the gas in the central parsec region is smaller.
This may be due to a reduction in the fueling rate as the
surrounding gas accretes onto the central engine \citep{bd02,ce02}. If
this scenario is correct, then the mean black hole masses in BL Lac objects
should be even greater than in FSRQs.  Consequently, the maximum
temperature of the disk in a BL Lac object should be
significantly lower than in FSRQs (recall that $\e_{max}\propto
M_{bh}^{-1/4} (\dot Mc^2/ L_{\rm Edd})^{1/4}$, where $L_{\rm Edd}$ is
the Eddington luminosity and $\dot M$ is the mass accretion rate; see
Dermer \& Schlickeiser 2002). In our calculations, we take 
$\e_{max} = 3\, \rm eV$for BL Lac objects.

We consider Mrk 501, a high-frequency peaked BL Lac object that has
been observed at TeV energies, as representative of this class of
sources. Mrk 501 has a dshift $z = 0.034$, so that $d_L = 160$ Mpc.
Variability has been detected on time scales as short as 1 hour,
although variability on time scales of order 1 day are also common. We
therefore consider both $t_{var} = 0.1$ and 1 day \citep{qui99,sam00}. 
For the compact jet
with $t_{var} = 0.1$ d, we approximate the flux density using the
April 16th, 1997 flare data \citep{pia98}, with indices
$\alpha_{1}\cong 0.5$ at frequencies $\nu_0 \cong 10^{11}{\rm ~Hz}<
\nu=\e/h \leq \nu_1 =10^{18}{\rm ~Hz}$, and $\alpha_{2}\cong
1$ at $\nu_1 \leq \nu \leq \nu_2 =10^{20}\,\rm Hz$, and assuming a very 
steep cutoff above $\nu_2$. The maximum $\nu F_\nu$
flux is at the level of $\sim 6\times 10^{-10}$ ergs cm$^{-2}$
s$^{-1}$. For the `quiescent' state with $t_{var}$ = 1 day, which may
correspond to a less powerful but continuous jet, we approximate the
spectral fluxes with the same power-law indices $\alpha$, but with
break frequencies $\nu_1 = 3\times 10^{14}$ Hz and $\nu_2 = 3\times
10^{17}$ Hz, and a maximum $\nu F_\nu$ flux at the level of $3\times
10^{-11}$ ergs cm$^{-2}$ s$^{-1}$.

Figure 6 shows the total energy spectrum of injected protons and
protons remaining in the blob at the time when the jet is at a
distance $R = 0.3$ pc, which we assume to be the effective radius of
the region where there is an isotropic scattered radiation field. In
this calculation, we treat the $t_{var} = 1$ d case, and let $\delta =
10$.  This results in an equipartition magnetic field 
$B_{eq} \cong 0.37\,\rm G$, and a maximum energy of accelerated protons 
equal to $9\times 10^{17}$\,eV in the comoving frame, which is larger
by a factor $\Gamma \sim 10$ in the stationary frame.  
For the external field, we assume that the maximum flux of
the disk radiation cannot exceed the flux of $3\times 10^{-11}$ ergs
cm$^{-2}$ s$^{-1}$ observed in the optical/UV range. The intensity of
the isotropic radiation field is calculated assuming $\tau_{\rm T} =
0.01$.  For these parameters of external radiation-field, there is
essentially no photomeson production by neutrons outside the blob, and
consequently very low gamma-ray production outside the blob
(dot-dashed curve).

The production of high-energy neutrons inside the blob through
photomeson interactions with the internal synchrotron photons is still
very effective for the highest energy protons with $E_p \gtrsim 10^{17}\,$eV. 
The fraction $W_n(>10^{17}{\rm eV})/W_p(>10^{15}{\rm eV}) = 0.12$, and 
practically all of this neutron energy is carried from the supposed isotropic
external radoiation to kpc scales.  The 
maximum energy of protons is, however, smaller compared with the 
$\delta = 10$ case for 3C 279 (Figure 1), which
implies a shorter propagation distance for the ultra-high energy (UHE)
neutron beam. The power deposited in the neutron beam during the flare
is $\gtrsim 4$ orders of magnitude smaller than the power of neutrons
escaping the BLR in Figure 1. Moreover, for larger Doppler factors, 
$\delta \sim 15$-45, which are implied by the modeling of TeV and X-ray
flares \citep{pia98,kca02}, the power in neutrons would decrease further. 
Our calculations show that $W_n(>10^{17}{\rm eV})\approx 2.6\times 10^{46}$ 
and $3.6\times 10^{44}$ ergs for $\delta = 10$ and 25, respectively.
Note that the fraction of energy carried by $E_\gamma > 10^{15}\,\rm eV$
gamma-rays produced by neutrons escaping the blob in the isotropic external
radiation region in Figure 1 is absolutely negligible,  
 $W_{\gamma}(>10^{15}{\rm eV})/W_p(>10^{15}{\rm eV}) \approx 4.7 
\times 10^{-5} $

In Figure 7, we show the fluxes calculated for the case $t_{var} =
0.1\,$d, assuming $\delta = 25$.  Figure 7a is calculated for $B^\prime
= B_{eq} = 0.5\,$G, corresponding to equipartition as defined in
equation (\ref{Beq}), whereas Figure 7b assumes $B^\prime = 0.2\,
B_{eq} = 0.1\,\rm G$.  The latter case may be of interest because the
spectral fits for the 16 April 1997 flare in Mrk 501 seem to require
fields well below equipartition. The maximum energies of accelerated
protons in the stationary frame are then $E_{max} \cong 7.5\times
10^{18}$ eV and $E_{max} = 1.5\times 10^{18}$\,GeV in Figures 7a and
7b, respectively, so that the effective maximum energy of the neutrons
escaping the inner jet (and therefore also the weak external radiation
field) does not significantly exceed $10^{18}\,\rm eV$, as seen in
Figure 7.  This suggests that in BL Lac objects, the transport of energy
by ultra relativistic neutrons is still possible, but the maximum
distances for energy transport by the neutrons will be significantly
shorter than in FSRQs.  Note also that the number of neutrinos to be
expected by a km-scale neutrino detector in these two cases is
extremely low, being less than $10^{-6}$ in both Figures 7a and 7b.

\section{Discussion and Summary}

Observations of powerful gamma-ray flares from blazars with EGRET and
ground-based gamma-ray telescopes have confirmed that radio-loud AGNs
accelerate particles to GeV and TeV energies, respectively. Particle
acceleration to higher energies may occur, but this cannot be
established through gamma-ray observations because of attenuation by
ambient and diffuse extragalactic optical and infrared radiation
fields.  Although leptonic models have been successful in fitting the
spectra of blazars, models of first-order Fermi acceleration at both 
nonrelativistic and relativistic shocks imply that acceleration
proceeds more effectively for protons than for electrons, because the
Larmor radius of a particle involved into the acceleration process
must be larger than the shock width, which is more easily satisfied by
hadrons \citep{gal02}. The large proton-to-electron ratio in the
galactic cosmic rays supports this contention. Thus our underlying
assumption that protons are injected with at least the same power as
inferred from the measured gamma-ray luminosity of a blazar seems
reasonable.

Evidence for the acceleration of relativistic protons and heavier
nuclei in blazar jets can be provided by three lines of evidence as
described in the following subsections.  First, hadronic acceleration
could be established indirectly by detection of spectral features in
the electromagnetic radiation produced by secondaries of the inelastic
interactions of high-energy hadrons in blazar jets. The second and
most compelling line of evidence will be provided through the direct
detection of high-energy neutrinos by km-scale detectors such as
IceCube.\footnote{http://icecube.wisc.edu/} The third line of evidence
involves consequences of the production of UHE neutrons and gamma-rays
which transport significant amounts of energy from the central region
of an AGN to large distances \citep{ad01}. This last feature is
sufficiently interesting that we devote a separate subsection to
applying this model to explain observed features of radio-loud AGNs.
Finally we summarize, considering also the possibility that ultra-high
energy cosmic rays (UHECRs) are accelerated in the inner jets of
blazars.

\subsection{Electromagnetic Radiation from Nonthermal Hadrons}

Discriminating between hadronic and leptonic origins of the gamma
radiation from blazars or other classes of sources is possible,
although not an easy or a straightforward task.  If the emission has a
hadronic origin, then the emerging gamma-ray spectrum results from a
cascade induced by high-energy secondary leptons and
gamma-rays. Photon spectral indices from cascade radiation tend to be
rather hard, and take values between 1.5 and 2.0 in the medium-energy
gamma-ray regime explored by EGRET and {\it
GLAST} \footnote{http://glast.gsfc.nasa.gov} (see Figure 5). But this is
also a common feature for gamma-ray spectra from an electron-photon
cascade independently of whether the cascade is initiated by
electrons and gamma-rays of secondary (pion-decay) origin or by
primary (directly-accelerated) electrons.

A principal difference between the hadronic and leptonic induced
cascades does, however, exist. It consists in a very significant
contribution of the synchrotron radiation by the ultra-relativistic
electrons (including the ones from the absorption of $\pi^0$-decay
gamma-rays) of the first cascade generation in the high energy
gamma-ray flux in the case of hadronic initiation of the cascade. The
Lorentz factor of these electrons $\gamma \gg 10^8$ is much higher
than the maximum possible Lorentz factor for primary electrons, which
is limited by synchrotron losses in the strong magnetic fields
$B\gtrsim 0.1$-$10\,\rm G$ characteristic for the inner jets of
blazars.  As a result, the synchrotron radiation of these electrons can
greatly exceed the characteristic maximum energy $E_{\rm s,max}^\prime
\sim 25$-100  MeV \citep{gfr83,jag96} for synchrotron emission of 
directly accelerated electrons, as is apparent from
Figure 5. Additionally taking into account the Doppler boosting of
the energy $E_{\rm s,max}$ when transforming to the observer's frame,
we can say that confirmation of a significant
contribution of hard-spectrum synchrotron flux in the gamma-ray flares
at $\gtrsim 0.1\,\rm GeV$ with {\it GLAST} could become a strong
argument in favor of UHE hadron acceleration in the jets.

A synchrotron contribution could be confirmed experimentally by the
detection of a significant polarization in the total flux, but neither
GLAST nor other high energy gamma-ray detectors yet have the
capability for the relevant measurements.  It is possible that the
synchrotron origin of the GeV radiation could be revealed by accurate
modeling of the spectral and temporal behaviors of the broad-band
gamma ray fluxes detected by GLAST, as well as by forthcoming
ground-based Cherenkov detectors. Of particular interest in this
regard could be the interpretation of flares which would {\it decline}
rapidly at 0.1-1 GeV energies at the flare fading stage. The key 
observation here is that in
FSRQs, the Compton radiation at these energies should be due to
upscattering of external UV photons by electrons with $E^\prime
\lesssim 1\,\rm GeV$, which could have cooling times larger than the 
measured decline time, whereas the GeV synchrotron radiation
is due to electrons with many orders of magnitude higher energies, and
so it would have much shorter cooling times.  

If radio galaxies are misdirected blazars and steep spectrum radio
sources are off-axis FSRQs \citep{bbr84,up95,ob82}, then the gamma-ray
spectra of radio galaxies could also reveal hadronic acceleration in
blazar jets.  Nonthermal cosmic rays that escape from the inner jets
of radio galaxies can accumulate in the central gas-rich region of the
AGN. Nuclear collisions of these cosmic rays with ambient gas would
produce a quasi-stationary (on time scales of thousands of years) and
compact gamma-ray halo with a characteristic $\pi^0$-decay feature
near 70/(1+z) MeV. Note however that the bremsstrahlung emission from
$\pi^\pm$-decay electrons of $\sim$0.1-1 GeV energies is likely to
conceal the 70 MeV benchmark \citep{sch82}.  Nuclear interactions will
give secondary $\pi$-decay emission that is brightest in the central
BLR core of a radio galaxy where the target particle density is highest, and
dimmer in the extended region due to the tenuousness of the
interstellar medium on scales of hundreds of pc.  Because of the
rather small size of this hadronic halo, the gamma-ray detectors would
detect it as a point source.  Such a compact hadronic halo could be
still resolved in the radio band, where the radio flux produced by
secondary electrons of GeV energies would show an increase in
brightness towards the center of the AGN on the milliarcsecond scale.
Note that at very high energies, a halo on the much larger scale 
(multi-Megaparsec) could be formed around such sources due to absorption 
and cascading of multi-TeV gamma-rays in the intergalactic medium 
\citep{acv94}.

Considering possible signatures of relativistic hadrons during the
flares, we note that a pronounced 70 MeV $\pi^0$-decay maximum  should
be due first of all to hadrons with kinetic energies 
$\sim 1$-10\,GeV; therefore
it would imply a mass-loaded jet. As we show in the Appendix A, the
mass-loaded jet models are rather unlikely because they confront severe
energetics difficulties. Nevertheless, the hadronic $\pi^0$-decay
gamma-ray feature, although significantly broadened and
Doppler-shifted, could be expected in those particular cases when the
inner jet dumps its energy in collisions with the dense target,
such as a broad line cloud crossing the jet.  Observations of radio
galaxies with imaging X-ray telescopes, {\it GLAST}, and ground-based
air Cherenkov telescopes with low energy thresholds will provide
spectral data that should be analyzed for evidence of nonthermal
hadrons in light of these considerations.

\subsection{Neutrinos from FSRQs and BL Lacs}

Detection of neutrinos from blazars will provide the most compelling
evidence for hadronic acceleration in blazar jets.  For photohadronic
jet models, we have argued here and elsewhere \citep{ad01} that the
presence of a strong UV accretion-disk radiation field is required for
blazar neutrino fluxes to be detectable with km-scale high-energy
neutrino telescopes. The disk radiation field is needed to produce a
quasi-isotropic scattered radiation component produced in the BLR of
FSRQs. The existence of an external radiation field in FSRQs is
suggested by luminous broad emission lines detected from these
objects. For UV fields expected from optically-thick accretion disks,
photomeson interactions can efficiently extract the energy of
relativistic protons with (relatively modest) energies $\gtrsim
10^{15} \,\rm eV$, resulting in the production of neutrinos with
$E_\nu \gtrsim 3\times 10^{13}\,\rm eV$. For flares like the one
detected in February 1996 from 3C 279 with EGRET, the fluence of
neutrinos shown in Figure 3 is at the level of $10^{-4} \;\rm erg\,
cm^{-2}$, implying $\approx0.3\; \nu_{\mu}$ per flare (i.e., a
detection probability of $\approx 30\%$).  This would suggest a
realistic possibility for IceCube to detect a few to several neutrinos
from several flares over the course of a year. Given the flaring duty
cycle of FSRQs estimated from EGRET observations ($\sim 10$-20\%),
this implies that a few high-energy neutrinos will be detected from
FSRQs such as 3C 279 on a time scale of 1 year. Considerably more will
be detected if the efficiency to accelerate hadrons exceeds our
baseline assumption given by equating proton and gamma-ray power.  The
additional possibility that some neutrinos will also be detected
during quiescent states strengthens our prediction that FSRQs will be
detectable high-energy neutrino sources with km-scale neutrino
telescopes.  We note that on a 1 year observation time scale,
detection of only two neutrinos with energies above 30 TeV from the
same direction pointed to a known blazar would represent a very
significant detection of an astrophysical source, because at these
energies the background cosmic-ray induced neutrino flux expected
within one square degree (which is typical for the angular resolution
of a high-energy neutrino telescope in ice or water) is $\lesssim
0.03$ \citep{ghs95}.

For BL Lac objects, where the accretion-disk and scattered radiation
components are much weaker, and with GeV gamma-ray fluxes that are
also typically not as bright as in FSRQs, the chances for detection of
neutrinos even with a 1 km-scale neutrino detector are negligible. For
the proton fluxes shown in Figures~6 and 7 calculated for parameters of
the April 1997 flare in Mrk 501, the probability of neutrino detection
during the flare is less than $10^{-6}$.  It is also important to note
that unlike the case of $p\gamma$ interactions with the internal
synchrotron photons, the intensity of $p\gamma$ interactions with the
external radiation does not depend on the blob size. Therefore in FSRQs 
the neutrino production from photomeson interactions with the external
radiation field does not decline so much in the process of expansion
of the blob as it propagates through the quasi-isotropic external
radiation field at $R\lesssim R_{\rm BLR}$ as it does in BL Lac
objects.  We note that the blob expansion seems unavoidable because of the
difficulty in confining the high internal pressures of the blob by the
pressure of the external medium.

An important issue concerns the quenching of the jet as it sweeps up
mass from the BLR.  As BLR clouds execute Keplerian rotational motion
about the central supermassive black hole, some will pass through the
path of the jet outflow.  This will cause energization and
deceleration of the radiating plasma, as in the external shock model
of gamma-ray bursts \citep{mr93}. Suppose that the AGN jet is beamed
into a fraction $f_b =0.01 f_{-2} $ of the full sky. In a
simple model with a uniform BLR gas density $n_{BLR}$ within a
spherical BLR with Thomson depth $\tau_{\rm T} = n_{BLR}\sigma_{\rm T}
R_{BLR}$, the total mass in the beaming cone is $M_j(<R_{BLR}) \cong
5250 f_b \tau_{\rm T} R_{18}^2 M_\odot$, where $R_{18} =
R_{BLR}/10^{18}\; {\rm cm}$ (see \citet{bl95} for BLRs with radially
dependent densities).  The time for BLR material to pass through the
cone of the jet with opening half-angle $\theta_j \cong 2\sqrt{f_b}$
is $t_j \cong 1.1\times 10^9 f_{-2}^{1/2} R_{18}^{3/2}/M_9^{1/2}$ s,
where the mass of the central supermassive black hole is $M_9 = M_{bh}
/10^9 M_\odot$. Assuming that the directional power of relativistic plasma
expelled in the AGN jets is comparable to the apparent isotropic
gamma-ray luminosity $L_\gamma$, then the flow will be significantly
decelerated when $L_\gamma t_j \cong \Gamma^2 M_j(<R_{BLR})c^2$, where
$\Gamma$ is the Lorentz factor of the plasma outflow. Thus, unless the
mean apparent isotropic luminosity exceeds
\begin{equation}
L_{\gamma}({\rm ergs~s}^{-1}) \; \cong \; 10^{48}\;
(f_{-2} R_{18} M_9)^{1/2}\;\left({\tau_{T}\over 0.1}\right)\,
\left({ \Gamma\over 10}\right)^2\;,
\end{equation} 
the jet will be quenched  by sweeping up BLR
gas. This quenching process could make it difficult to form extended
relativistic jets on the 10-100 kpc scale, as may be required to
explain enhanced X-ray emission in the knots of PKS 0637-752
\citep{tav00}. The transport of inner jet energy to the extended jets
via neutral beams avoids this difficulty. The neutrons and gamma-rays
will pass through the BLR material without attenuation unless the
Thomson depth of BLR is very high, with $\tau_{T} \gtrsim 10$. In these
cases the jet and neutral beam outflow would be quenched, the
nonthermal radiation would be reprocessed into lower energy emissions,
and only the neutrino signal would penetrate through the dense inner
region without reprocessing. If so, the neutrino emission from such
buried jets in non-blazar sources could reveal these types of AGN. We
return to this speculation in Section 4.4.

\subsection{Energy Transport to the Extended Jet by Neutral Beams} 

Relativistic neutrons produced by $p\gamma$ collisions in the compact
inner jet are no longer confined by the jet's magnetic field, and so
can escape from the blob unless the neutron decays or interacts with
another soft photon that results in conversion to a proton inside the
blob. Our calculations show that in FSRQs with luminous external
radiation fields, up to $\sim 30\%$ of the total energy injected into
the blob in the form of relativistic protons with $E\gtrsim
10^{15}\,\rm eV$ can leave the blob in the form of relativistic neutrons
 with Lorentz factors $\gamma_n \gtrsim 10^{6}$.

For the case of FSRQs, neutrons escaping the inner blob continue to
propagate in the BLR with its quasi-isotropic scattered accretion-disk
radiation up to distances $R\sim R_{BLR} \sim 0.1$-1 pc. Subsequent
interactions of the neutron beam with the scattered disk photons will
result in the production of beamed secondaries.  Neutrino secondaries
will add to the overall neutrino flux. Importantly, a beam of
gamma-rays with energies above $10^{15} \,\rm eV$ is also produced.
Because these gamma-rays are made outside the blob and propagate radially
outward in the blob's frame, they do not
interact with the internal synchrotron photons, and therefore have a
chance to escape from the central region of the quasar (see Figure 
2). Depending on the extent and intensity of the isotropic external
radiation field, as much as several per cent of the accelerated proton
energy $W_{p}(\geq 10^{15} \,\rm eV$) injected in the inner jet at
distances $R\leq R_{BLR} $ could thus be converted to gamma-rays which
escape from the BLR.

Figure 8 shows the mean-free-path of gamma rays due to pair-production
interactions with photons of the diffuse cosmic microwave background
radiation for $z = 0$ (solid curve) and $z = 0.538$ (dashed
curve). This calculation also takes into account the diffuse
extragalactic infrared radiation field.  We approximate this field by
a spectrum $\nu J_\nu \propto const$ at the level of total energy
density $u_{IR} = 0.01$ eV cm$^{-3}$ between $0.001$\,mm and 1\,mm
\citep{pri99}, though the precise value is not important on the size
scales $\lesssim 1$ Mpc under consideration.  

Our calculations show that more energy is typically carried out from the
BLR by relativistic neutrons than by gamma rays. This is especially
true if the degradation of the neutron beam escaping the blob but still
in the BLR is not
very strong, which will always be the case for neutrons escaping the
blob at distances $R \sim R_{BLR}$. The energy carried away by
neutrons with $E_n \geq 10^{17} \,\rm eV$ may reach $\sim 10\%$ of
$W_{p}(\geq 10^{15} \,\rm eV$) for accelerated protons with number
index $\alpha_{p}\simeq 2$. Adding the energy of the gamma rays to
this, we find that $\sim 5\%$ can be used as a rough estimate for
the fraction of total (starting from 1 GeV) energy of accelerated protons 
in the inner jet
that is carried by the collimated beam of UHE neutrons and gamma rays
 to $>$kpc distances in FSRQs.

For a flare like the one in February 1996 from 3C 279, the gamma-ray
fluence measured with EGRET corresponds to an apparent isotropic
energy release of $2\times 10^{54}\,\rm erg$. Assuming that the
injection power in nonthermal protons in the inner jet is the same as
for the parent electrons, the overall total energy in the neutral beam
emerging from the inner jet per a single such flare could reach the
level $\sim 10^{51} \delta_{10}^{-2} \,\rm erg$ in the stationary
frame. This energy will be collimated in the cone with opening 
angle $\sim
1/\delta$, and it will be deposited at distances up to 0.1-1
Mpc, provided that the spectra of ultra-high energy neutrons extend
up to $10^{19}$-$10^{20}\,\rm eV$, as in Figure 4b.

After neutron-decay or gamma-ray attenuation, the charged particles
emerging in the direction of the parent neutral beam will interact
with the ambient extragalactic medium via local magnetic and photon
fields.  In the case of the neutrons, a small fraction $\sim m_e/m_p$
of the neutron energy will appear in the form of $\beta$-decay
electrons with Lorentz-factors $\gamma > 10^8$.  This energy will be
immediately available for radiation in the synchrotron and Compton
processes with $\nu F_\nu$ peaks at hard X-ray and multi-TeV
energies, respectively, for a 1 $\mu$G field and a dominant local CMB
radiation field \citep{der02}.  The dominant fraction of the initial
neutron energy remains, however, in the protons.

 Synchrotron radiation of ultrarelativistic neutron-decay protons 
could in principle explain the extended {\it Chandra} X-ray jet emission
if the ambient magnetic field is sufficiently large \citep{aha02}.
Proton synchrotron radiation does not, however, provide a mechanism 
for deposition of energy of the neutral beam into the surrounding medium
to drive shocks and accelerate  $\sim 1$-100 GeV electrons, which is 
required to explain the radio and optical emission from the extended jets.

In order for the transport of energy 
by the neutrons to be efficient from the point of view of
 powering the extended jets in radio galaxies, there should be
a mechanism for effective transfer of the neutron-decay proton energy
 to the surrounding plasma.
Such a mechanism is expected through interactions of the relativistic
protons as well as the $e^{+}$-$e^{-}$ pairs from the gamma-ray beam
with the ambient magnetic field in the surrounding medium.  In the
case of a significant transverse magnetic-field component $B_\perp$,
both the protons and leptons will change their initial directions on
the Larmor timescales $t_L = E/eB_\perp c$, which implies that a
significant fraction of their momentum and energy will be transferred to
the ambient medium.  Another, and possibly even more effective way for
transferring the initial momentum and energy of the charged particles
is by generating MHD turbulence through beam instabilities excited by
the highly beamed neutron-decay protons and charged secondaries
emerging from the charge-neutral beam.  The continuous transfer of
momentum and energy from the beam to the intergalactic medium could
stretch the ambient magnetic field to form a channel facilitating beam
propagation, or drive the medium into relativistic motion in the
forward direction to create a powerful relativistic shock.

In the latter scenario, relativistic shocks can accelerate ambient
electrons (as well as protons), thereby injecting a power-law electron
spectra $Q_{e}(E) \propto E^{-\alpha} $ with $\alpha \geq 2$ in the
downstream region.  Synchrotron radiation from these electrons could
then explain \citep{da02} the broad-band nonthermal radio, optical,
and X-ray emission from hot spots and knots in the extended X-ray jets
of extragalactic jet sources observed with the {\it Chandra
Observatory}.  The idea that the observed X-ray jets might be due to
propagation of a powerful beam of gamma rays has also been considered
by \citet{nsak02} \citep{dl01}, who suggested a
``non-acceleration" scenario, where the observed X-rays could be due
to synchrotron radiation from electrons formed in the electromagnetic
cascade initiated by ultra-high energy gamma rays. Our calculations
show that although some contribution to the observed X-ray fluxes from
the cascade electrons is not excluded, the spectral features of the
observed radiation, in particular, X-ray energy spectral indices
$\alpha_X \geq 1$, might be difficult to explain in a pure cascade
scenario. Moreover, interpretation of the radio and optical fluxes
from the same X-ray knots apparently makes unavoidable the requirement
of in-situ acceleration of electrons in the X-ray knots.  Given the
total energy released by the neutrons and gamma rays per single flare,
the energetics of these kpc-scale knots could be explained by
superposition of jet activity on time scales up to thousands of yrs
(since $1 \,\rm kpc$ translates to $\simeq 3\times 10^3 \,\rm yr$ 
of light travel time).  Thus, the
bright knots might represent the observable signatures of the past
history of activity of the central engine of the blazar when powerful
acceleration of UHE cosmic rays had occurred in the inner jets
of the source.

Considering now the BL Lac objects, we note that in the case of weak
or negligible external radiation field components, any significant
transport of the inner jet energy by ultra-high energy gamma-rays in
those objects is practically absent. This is because gamma-rays are
not produced through $n\gamma$ interactions of the neutrons outside
the compact blob, whereas the ultra-high energy gamma-rays produced in
$p\gamma$ collisions inside the blob will be absorbed due to
$\gamma\gamma$ collisions with the same internal synchrotron target
photons (recall that $\sigma_{\gamma\gamma} \gg \sigma_{p\gamma}$).
Nevertheless, up to $\sim 10\%$ of the energy injected into the inner
jet in relativistic protons with $E>1\, \rm PeV$ could still be taken
out of the blob by neutrons with $E\gtrsim 10^{17} \,\rm eV$ due to
$p\gamma$ collisions with synchrotron photons while the blob
remains compact. Because of the latter limitations, and because BL Lac
sources are generally much less powerful than FSRQs, the total energy
that the neutron beam can extract from the accelerated protons is much
less for BL Lacs than for FSRQs. Moreover, because of the smaller sizes
and magnetic fields inferred for BL Lac objects, the level
of neutron production and the maximum energy of accelerated protons is
reduced in the former class of sources. Thus the neutral beam power is
much less in BL Lac objects than in FSRQs.

\subsection{Explanation of Observed Features of Radio-Loud Active Galactic
 Nuclei }

As a consequence of processes occurring in the inner jets of blazars,
we predict that powerful collimated neutral beams that extend as
far as hundreds of kpc from the nucleus are ejected in the axial jet
directions in high luminosity FSRQs. The linearity and the spatial 
extent of the ejecta is a
consequence of the decay properties of the neutrals.  This process can
explain the origin of X-ray jets detected by {\it Chandra}, which
appear straight on scales of up to Mpc, such as the jet in Pictor A
\citep{wys01}.  The linearity of the radio jets in Cygnus A
\citep{pdc84} may also be a consequence of energetic neutral beams
powering the lobes of powerful FR II galaxies. This interpretation
avoids firehose instabilities in models invoking magnetic fields to
collimate the extended FR II radio jets, or problems associated with
jet quenching, as discussed in Section 4.2.

Our model predicts that lower luminosity blazars, in particular, the
lineless BL Lac objects, have orders-of-magnitude weaker neutral beams
than FSRQs. The maximum energies of the relativistic neutrons are also
about an order-of-magnitude less in Mrk 501-like BL Lac objects than
in 3C 279-like FSRQs. 
Consequently the X-ray jets driven by the high-energy neutron
beams will be generally fainter (despite their relative proximity) and
shorter in BL Lac objects. This could explain the faint kpc-scale
X-ray jets detected from objects like 3C 371 \citep{3C371}, which is
an off-axis BL Lac object \citep{mil75}.

We adopt the point of view, following \citet{up95}, that FR I and FR
II galaxies are the parent populations of BL Lac objects and FSRQs,
respectively.  The less powerful beams of neutrons that occupy smaller
spatial scales in BL Lac objects compared to FSRQs conforms to the
measured relative power requirements in FR I and FR II galaxies
\citep{fr74}. If FR I/BL Lacs have smaller Lorentz factors than FR
II/FSRQs, as also inferred from observations of these sources
\citep{up95}, then the weak, broad twin-jet morphologies in FR I
galaxies can be qualitatively understood.  We point out that even
though the inner parsec regions of a BL Lac are much less dense than
the BLRs in FSRQs in view of their relative emission line strengths,
the jet quenching problem (Section 4.2) can still be severe for BL
Lac objects because of their weaker jets.

The transport of energy by neutral particles to the extended jet will
be enhanced by an intense external radiation field in the inner
regions of the jet.  The total power into the jet depends foremost on
the accretion rate, which depends on the amount of gas and dust
available to fuel a supermassive black hole. Thus our model is
consistent with the evolutionary scenario proposed by \citet{bd02} and
\citet{ce02} to explain the spectral
sequence of FSRQs, and low- and high-frequency peaked BL Lac
objects. According to this picture, FSRQs evolve into BL Lac objects
due to a reduction in the accretion rate of the surrounding dust and
gas that fuels the central black hole. (Note, however, that the jet
power is derived primarily from mass accretion in the scenario of
\citet{bd02} and from the spin energy of the black hole in the
scenario of \citet{ce02}.) \citet{mt02} argue that this picture
accounts for the correlation between the radio jet and accretion
power, where the latter is inferred from emission line strengths.
High-energy neutrino observations will be crucial to test this
scenario, as neutral beams of high energy neutrons and gamma-rays will
be made in association with high-energy neutrinos.  Finally we
speculate that if ultraluminous infrared galaxies and quasars evolve
into radio-loud AGNs \citep{san88}, then some ultraluminous infrared
galaxies could harbor buried jet sources that will be revealed
through their high-energy neutrino emission.

\subsection{Summary}

Production of narrow beams of ultra-high energy neutrons with $E\geq
10^{17} \,\rm eV$ becomes possible if relativistic compact jets of
blazars accelerate cosmic rays to these energies. Such neutral
relativistic particle beams can be especially powerful in FSRQs, where
the neutrons are efficiently produced in $p\gamma$ interactions with
photons of the external quasi-isotropic radiation field on timescale
of years (since the broad-line region size scale $R_{BLR}\sim 0.1$-1
pc).  In radio-loud jet sources, the beam could contain a significant
fraction of ultra-high energy gamma-rays produced in the BLR through
$n\gamma$ photomeson interactions by neutrons which escape from the
compact relativistic plasma blob that forms the inner jet.  These
charge-neutral beams of UHE neutrons and gamma-rays (including also
neutrinos which, however, do not contribute to the process of energy
transport) do not interact with the extragalactic medium until either
the neutrons decay or the gamma rays collide with soft photons on
scales from kpc to Mpc. The main soft-photon target for $\gamma
\gamma$ collisions is provided by the CMB radiation and, in some
sufficiently powerful sources, also by low-frequency synchrotron radiation
self-produced along the jet 
\citep{nsak02}).

The weaker neutral beams in FR I radio galaxies and BL Lac objects,
compared to FR II radio galaxies and FSRQs, could explain the
morphological differences between the two classes of radio
galaxies. This scenario assumes that the inner jets of blazars are
effective accelerators of high-energy cosmic rays. The limits on
acceleration that we consider do not preclude the acceleration of
cosmic rays to energies $\gtrsim 10^{20}\,\rm eV$ in blazars,
especially in the more luminous FSRQs.  This mechanism of relativistic
energy transport from the central engine of AGN to large distances
avoids problems of jet quenching in the central gas-rich regions of
AGN, problems of adiabatic and radiative losses, and problems of
long-term confinement and stability (on scales up to $10^{6} \,\rm
yr$) of the jet against different plasma processes. Beam
instabilities can, however, provide an effective method to transfer
the energy of ultra-relativistic protons emerging from the neutron
beam into the surrounding medium, and driving relativistic shock which
ultimately accelerates nonthermal electrons to radiate observable
synchrotron and Compton emission. Given also the large amount of
energy that such a beam may contain, which can be of the order of a
few percent of the inner jet power, our results suggest that the
detection by the {\it Chandra Observatory} of the narrow X-ray jets
that remain straight on scales beyond several hundreds of kpc provides
observational evidence that the relativistic inner jets of blazars,
first of all the flat spectrum radio quasars, are powerful
accelerators of UHE cosmic rays. Detection of high-energy neutrinos
from FSRQs with IceCube or a Northern Hemisphere high-energy neutrino
telescope will provide strong support for this scenario.
   
\acknowledgments{We thank Hui Li for discussions about the physics
of jets, and the referee for a very useful report. AA appreciates the
hospitality and support of the NRL High Energy Space Environment
Branch during his visit when this work has been done.  The work of CD
is supported by the Office of Naval Research and NASA grant No. DPR
S-13756G.}

\appendix

\section{Comparison of Mass-loaded and Photopion Hadronic Jet Models} 

Relativistic neutrons are produced with $\sim 50\%$ probability in the
inelastic collisions of relativistic protons either with 
an ambient thermal proton or with a photon above the photopion production 
threshold. However, the mass-loaded jet models, i.e. the ones 
invoking $pp$ nuclear collisions, can be practically excluded 
on the basis of efficiency arguments. 
As equation (\ref{tpp}) shows, nuclear interactions
could provide energy loss time scales shorter than the dynamical
time $t_{dyn}^\prime = R^\prime/c$ only if the source is very thick to   
Thomson scattering, $\tau_{sc } > 40$. If one requires that, say, 
non-thermal X-ray flux are produced during the flare, i.e. $\tau_{sc} <1$,
then the observed fast decline of the fluxes  
could be attributed to the adiabatic losses of proton energy and fast 
decline of the gas density on dynamical time scale 
due to rapid expansion of the source. 
However, the overall energy losses 
of protons in nuclear interactions during that time cannot exceed 
$\simeq 2.5\%$. In practice this means that the real power injected into
the blob in the form of relativistic protons should be by a factor of
$\gtrsim 100$ larger  than the power inferred from the observed
radiation fluence.

Besides these inefficiency problems, a serious argument against
hadronic nuclear interaction models is that the mass-loaded jets would
require too great a total energy in the blob including the kinetic
energy. Thus, for a blob size $R^\prime \sim t_{var} c
\delta/(1+z)$ with typical $t_{var}\simeq 1 \,\rm d$ and $\delta
\simeq 10$ the Thomson scattering depth $\tau_{sc} =1$ implies a gas
density (in terms of 'H-atoms', and omitting here the dependence on
the source redshift $z$) $n_{\rm H} \sim 6 \times 10^7 \,\rm cm^{-3}$.
The jet mass then is $M_{j} \simeq 3.5
\,M_\odot$, 
and the kinetic energy $W_{kin} = (\Gamma -1) M_{j} c^2
\sim 6 \times 10^{55}\,$erg for $\Gamma \simeq \delta \sim 10$.  
Thus, even such a jet with reduced efficiency should have a
significant mass load leading to an unacceptably high energetics
required in a single blob. It would be formally possible to bring the
high kinetic energy requirements alone \citep{aha99} down to a
formally acceptable values by reducing further the density of the
gas. However, then the total energy of relativistic protons $W_p$ in
the blob should be increased in order to provide a given level of
radiation fluxes by $pp$-interactions.

Indeed, the integral luminosity of the radiation
that could be produced in $pp$-interactions, is given by 
$L_{r}^{\prime} \simeq \eta_r W_{p}^\prime /t_{pp}^{\prime}$, where
$\eta$ is the fraction of relativistic proton energy that goes to
secondary electrons and gamma-rays. Assuming that most of the electron 
energy will be processed to electromagnetic radiation, we can have 
$\eta \cong 0.5$,
leaving the rest for the neutrinos. Transforming from the blob frame to the
observer (or stationary) frame, the total energy $W_{tot} =
W_{p} + W_{kin} $ can be written as 
\begin{equation}
W_{\rm tot} \simeq 4\pi \Gamma m_p c^2 \left[
\frac{4 d^2 f_{\rm tot} (1+z)^4 }{ \sigma_{pp} m_p c^3 \delta^4 } 
(n_{\rm H}^\prime)^{-1} +
\frac{(t_{\rm var} c \delta )^3}{3 (1+z)^3} n_{\rm H}^\prime \right] \; ,
\label{Wtot}
\end{equation}
where $f_{\rm tot} $ is the total energy flux, i.e. 
$\int F(\nu) {\rm d} \nu$
 which is typically larger than the $\nu F(\nu)$ that a model is intended 
to explain. 
Here we have assumed for the kinetic energy 
$(\Gamma-1)/\Gamma\simeq 1$, and substituted $K_{pp} =0.5$ for the 
inelasticity.
The energy $W_{\rm tot}$ reaches a minimum at the density
\begin{equation}
n_{\rm H.min}^\prime \simeq 2.25 \times 10^9 \, d_{28} f_{-10}^{1/2}
\, t_{\rm var }^{-3/2}(d) \;  \delta^{-7/2}
\, (1+z)^{7/2}
 \; \rm cm^{-3} \; ,        
\label{nHmin}
\end{equation}
where $f_{-10} \equiv f_{\rm tot}/10^{-10} \,\rm erg \, cm^{-2} \, s^{-1}$,
and $d_{28} \equiv d /10^{28}\,\rm cm$. The minimum of 
$W_{\rm tot}$ is then 
\begin{equation}
W_{\rm min} \simeq 1.6 \times 10^{54}\;  d_{28} f_{-10}^{1/2} 
\, t_{\rm var}^{3/2}(d)
             \;    \delta_{10}^{1/2} (1+z)^{-1/2} \; \rm erg\, .
\label{Wmin}
\end{equation}
Here $\delta_{10} \equiv \delta /10$, and  we have assumed $\Gamma = \delta$
in equation (\ref{Wtot}). Interestingly, $W_{\rm min}$ is reached 
when the kinetic and internal CR energies become equal.
Note that any noticeable deviation of the real 
density in the jet from $n_{\rm H.min}$ required for $W_{\rm min}$ will
immediately increase further the total energy requirement given by equation 
(\ref{Wtot}).  

A hadronic nuclear interaction jet model which could make any
significant contribution to the fluxes of flares observed by EGRET
from FSRQs, would have to assume an
absolute (in the stationary frame) energy in a single flare exceeding
by orders of magnitude an energy of a supernova explosion. Even for a
less powerful but much closer blazars, like BL Lac objects Mrk 421 or
Mrk 501, any significant contribution to the total radiation flux from
mass-loaded nuclear interaction jet models can be practically
excluded. Indeed, even taking into account that the variability
timescale might be as short as 1 h instead of 1 day, for
interpretation of the maximum fluxes at the level of $f_{\epsilon}
\simeq 10^{-9} \,\rm erg\, cm^{-2} s^{-1}$ the minimum energy $W_{\rm
min}$ would still significantly exceed the `supernova' level $10^{51}
\,\rm erg$, which is problematic to accept for these objects with a
relatively modest average bolometric luminosities.  The only type of
relativistic extragalactic sources for which the hadronic nuclear
interaction models could still be reasonable from the general
total-energy consideration seem GRB sources where variability
timescale is very short (hence the densities could be very high)
although the peak fluxes are much higher than for relativistic
jets. Noticing that $f_{\rm tot} \times t_{var}$ is actually a good
estimate for the fluence $\Phi $ observed during the flare, we can
re-write equation (\ref{Wmin}) as $$ W_{\rm min} \simeq 6.3 \times
10^{49}\; d_{28} \Phi_{-5}^{1/2}
\, (t_{\rm var}/1 {\rm s}) \;    (\delta /100)^{1/2} \,
(1+z)^{-1/2}\;
 \rm erg\, ,
$$
where $\Phi_{-5} \equiv \Phi / 10^{-5} \,\rm erg\, cm^{-2}$. Obviously,
this is a reasonable value for GRB source, although the question on the
efficiency of the $pp$-interaction model in case of simultaneous detection 
of non-thermal X-rays and gamma-rays should be still addressed. 

\section{Relative Importance of the Direct Disk and Scattered Radiation 
Fields for Photomeson Production}

Equations (\ref{ss}) and (\ref{uext}) can now be used to estimate the
relative importance of the photomeson interactions of relativistic
nucleons with the direct and scattered components of the accretion
disk radiation, respectively. For the direct disk radiation
photon spectral density $n(\e,\Omega) = u(\e , \Omega; h )/\epsilon$,
we analytically estimate the dimensionless number of photomeson
collisions per unit path $dh$ (in terms of gravitational radius
$R_g$), $\kappa_{dir} = \nu_{p\gamma} R_g/c $, where the collision
rate $ \nu_{p\gamma} $ is given by an integral (\ref{tpgamma}) without
the inelasticity term $K_{p\gamma}$. This yields the simple formula
\begin{equation}
\kappa_{dir}(h,E) \approx 4.1 \times 10^3 \, l_{ad} \frac{I(h,E)}{h^{9/4}} 
\left( \frac{\epsilon_\ast}{100\,\rm eV} \right)^{-1} \frac{ \bar{\sigma}}
{ 200 \, \mu \rm b} \; .
\label{kappaDir}
\end{equation}
Here $l_{ad}=L_{ad}/L_{Edd} $ is the total luminosity of the disk radiation
in units of Eddington luminosity $L_{Edd}$, and 
\begin{equation}
I(h,E) = \int \frac{1- \cos \psi}{\tan^{9/4}\psi} d \cos \psi\; ,
\label{Int}
\end{equation}
where the integration is to be performed only over the collision angles
$\psi$ satisfying the threshold condition 
\begin{equation}
C(\psi)\equiv 
\frac{1-\cos\psi}{\tan^{3/4} \psi } \geq \frac{\epsilon_{th} h^{3/4}}
{\epsilon_{\ast} \gamma_{p,n} } \; ,
\label{threshold} 
\end{equation}
with $\gamma_{p,n}=E/m_{p}c^2$ (for $m_n = m_p$) corresponding to the
Lorenz-factor of the incident proton or neutron. For simplicity of the
analytic estimate (\ref{kappaDir}), we here approximate the $p\g$
interaction cross section $\sigma(\epsilon_{\rm r})$ as a single
step-function with a height $\bar\sigma \sim 200 \, \rm \mu b$ between
$\sigma_1 $ and $\sigma_2 $, starting from the threshold energy
$\epsilon_{th} \simeq 150 \,\rm MeV$.

A similar analytic estimate for the coefficient $\kappa_{\rm iso}(E)$ 
is derived 
for the scattered quasi-isotropic component of the external radiation field
(\ref{uext}). For distances $h\leq r_{BLR}$, i.e. inside the BLR with
dimensionless radius $r_{BLR}$, this results in
\begin{equation}     
\kappa_{\rm iso}(E) \simeq 1.1 \times 10^5 \;  
\frac{ l_{ad} \tau_{\rm T}}{ \, r_{BLR}^{2}}
\left(  \frac{\epsilon_{max}}{20\,\rm eV} \right)^{-1}
 \frac{ \bar\sigma}{ 200 \, \mu \rm b} \; ,  
\label{kappaIso}
\end{equation}
where $\epsilon_{max} = \epsilon_\ast/r_{i}^{3/4}$, with $r_i\geq 6$,
as described above.

Because the scattered radiation is quasi-isotropic, the interaction
threshold condition for photopion production in this field is the same
at all heights $h< r_{BLR}$, and is defined by head-on collisions of 
protons and neutrons with photons with maximum energy $\epsilon \simeq
\epsilon_{max}$. For $\epsilon_{max}
\simeq 20\,\rm eV$, this requires nucleon energies $E > E_{th}
\simeq 4 \times 10^6 \, m_{p}c^2$. Meanwhile, because of the
follow-on collisions that nucleons, which are moving predominantly
along the jet away from the black hole, make with the photons of the
direct disk radiation field, the interaction threshold of the nucleons
with this field significantly increases with increasing height
$h$. From equation (\ref{threshold}), we find $E > E_{dir}(h)
\simeq 4.5 \times 10^6 h^{3/4}m_pc^2 $, taking into account that the
function $C_{\psi}$ reaches a maximum $\cong 0.331$ at $\psi =
60^\circ$. This means that only particles with energies $E \gg 10^{17}
\,\rm eV$ have effective interactions with the direct disk
radiation component once the jet reaches heights $h\gtrsim 100
$. Meanwhile, interactions of {\it all} particles with energies above
4 PeV can effectively continue until heights $h \simeq r_{BLR} \sim
2\times 10^3$-$10^4$ (corresponding to $R_{BLR}\sim 0.1$-1\,pc for a
black hole mass $M_{bh} \sim 10^9 \,M_\odot$). 

The characteristic photomeson opacity $\tau_{\rm iso}
\simeq \kappa_{\rm iso} r_{BLR}$ due to interactions with
 BLR photons can be very significant for $l_{ad}\gtrsim 0.1$ and
$\tau_{\rm T} \sim 0.1$. For comparison, the characteristic photomeson
opacity $\tau_{dir} = \kappa_{dir}\, h$ from the direct disk radiation
field on scales $h$ can be significant ($\gtrsim 1$) only rather close
to the black hole, at $h < 100$. This applies only to particles well
above the interaction threshold $E_{dir}(h)$, when the integration
limits in the integral (\ref{Int}) cover a significant fraction of the
maximum allowed range $0 < \cos\psi < 1$, when $I(h,E)$ reaches a
maximum about $0.25$. Given these considerations, we consider only
interactions with the quasi-isotropic (i.e, scattered BLR field)
component of the external disk radiation. Note that the acceleration
of nonthermal particles may occur at large distances from the central
black hole if due either to collision of relativistic shells or to
interactions of the outflowing plasma with ambient gas in the BLR.

\clearpage
\begin{figure}
\epsscale{0.7}
\plotone{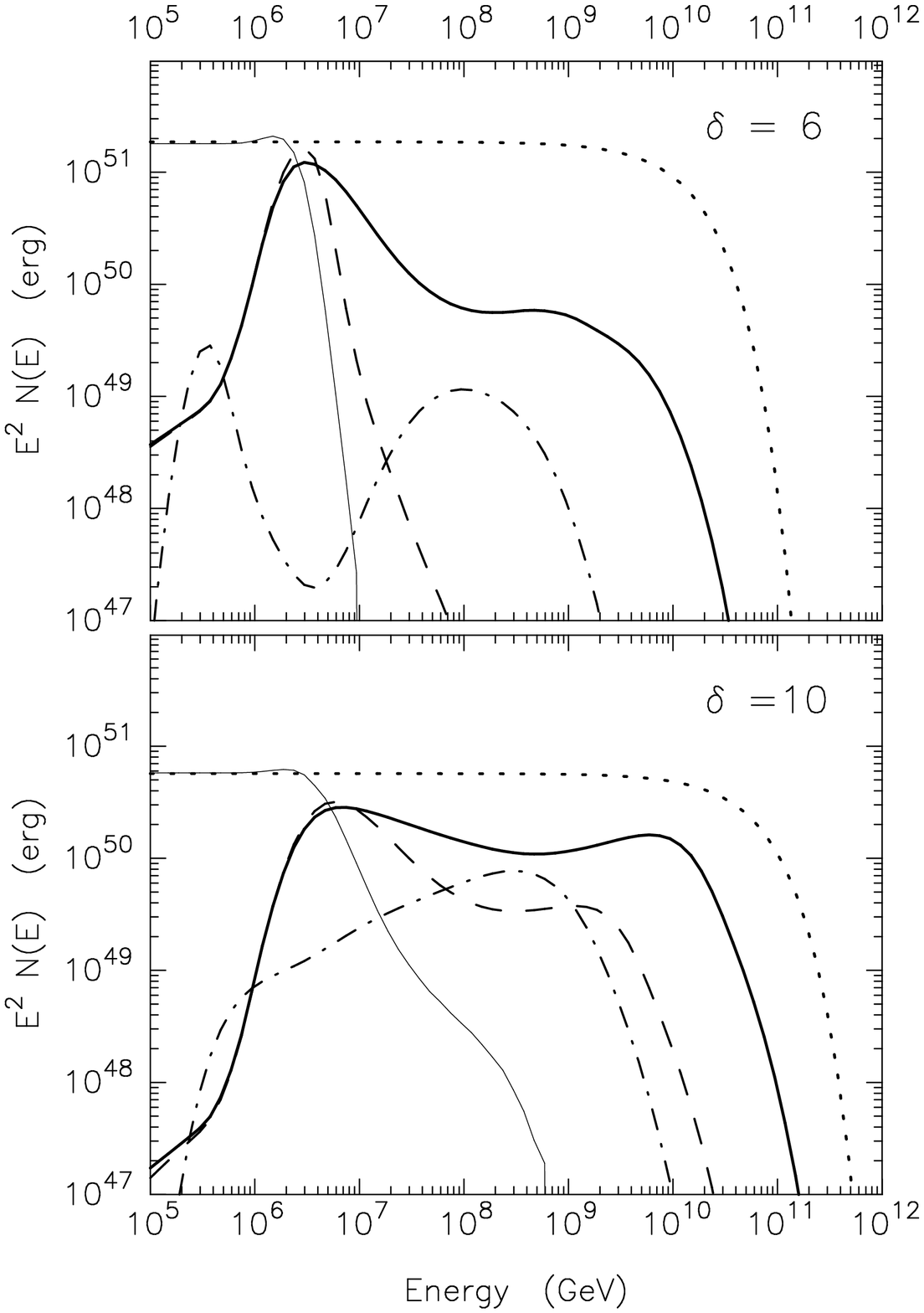}
\caption{Spectra of protons injected into the blob (dotted curves), protons
 which remain in the blob when it reaches the edge of the BLR (thin
 solid curves), neutrons escaping from the blob (thick solid curves),
 and escaping neutrons (dashed curves) and gamma rays (dot-dashed
 curves) which reach the edge of the BLR for the cases of $\delta = 6$
 and $\delta = 10$ in the upper panel (a) and lower panel (b),
 respectively. Model parameters relevant to the February 4-6 flare 
from 3C 279 detected by EGRET are used, as described in Section 3.1. }
\label{f1}
\end{figure}

\clearpage

\begin{figure}
\epsscale{0.9}
\plotone{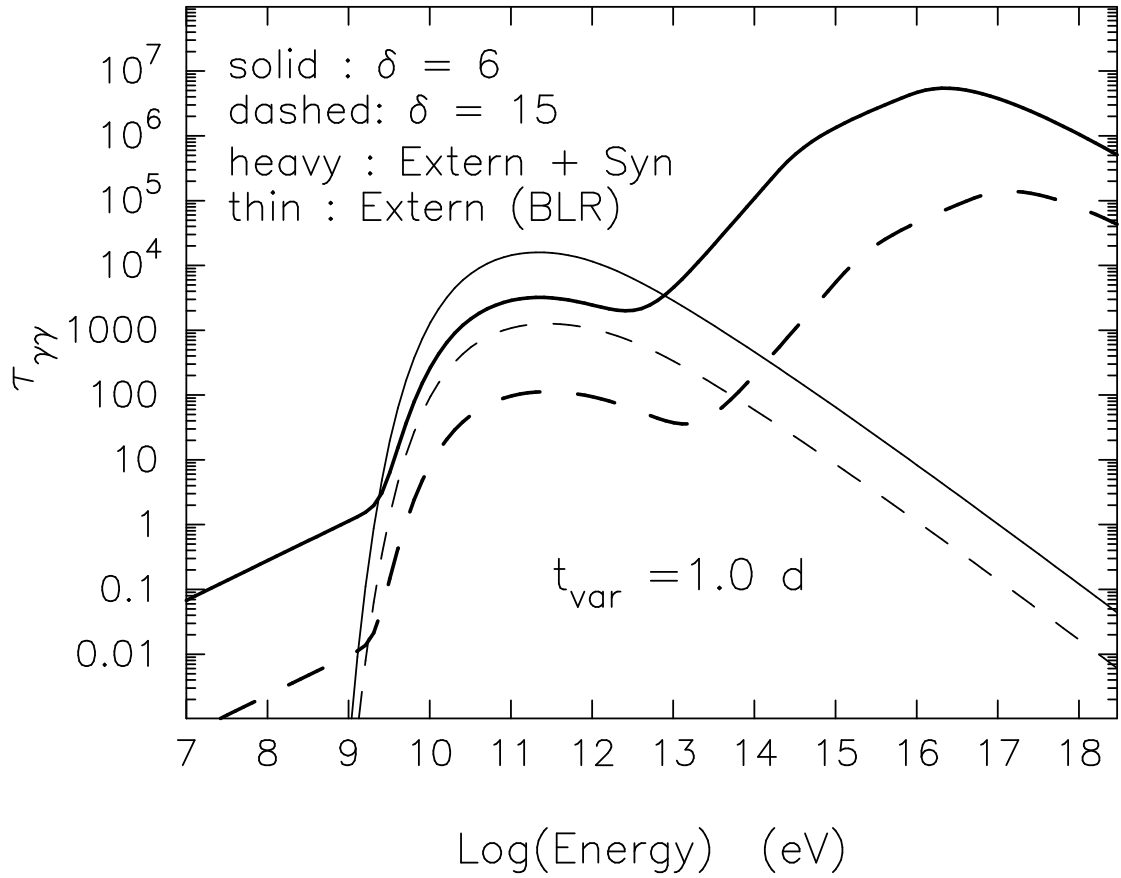}
\caption{Pair production opacity due to $\gamma\gamma$ interactions inside 
the blob and in the BLR, as indicated in the figure legend. }
\label{f2}
\end{figure}

\clearpage

\begin{figure}
\epsscale{1.0}
\plotone{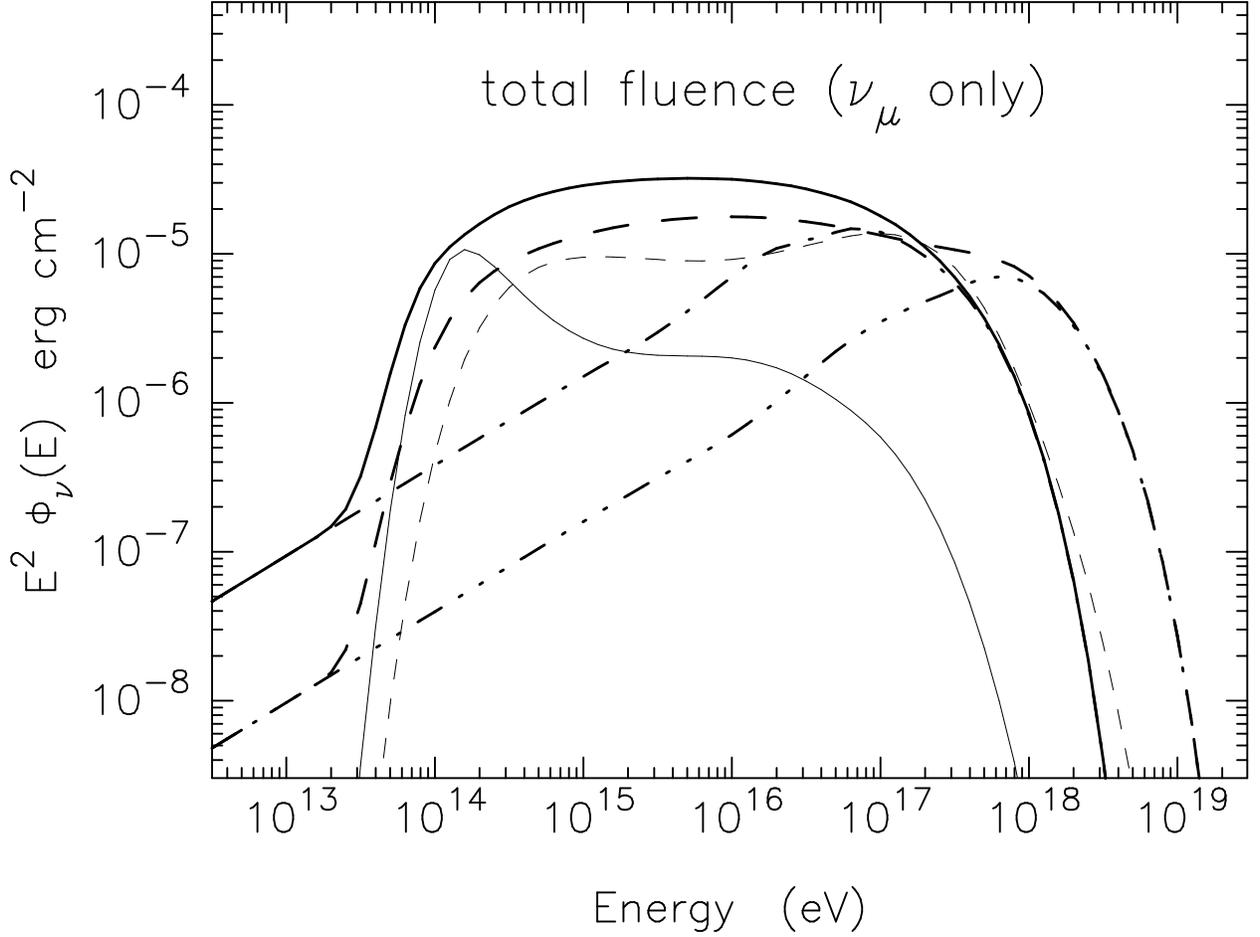}
\caption{Fluences of neutrinos integrated over several days
in the observer frame determined by the time for the blob to pass
through the BLR, for the same parameters as used in Figure 1. The solid
and dashed curves show the fluences calculated for $\delta = 6$ and
10, and the thick and thin curves represent the fluences of neutrinos
produced by photopion interactions inside and outside the blob,
respectively. The dot-dashed and 3-dot -- dashed curves show the fluences
due to $p\gamma$ collisions if external radiation field is not taken
into account.}
\label{f3}
\end{figure}

\clearpage

\begin{figure}
\epsscale{0.6}
\plotone{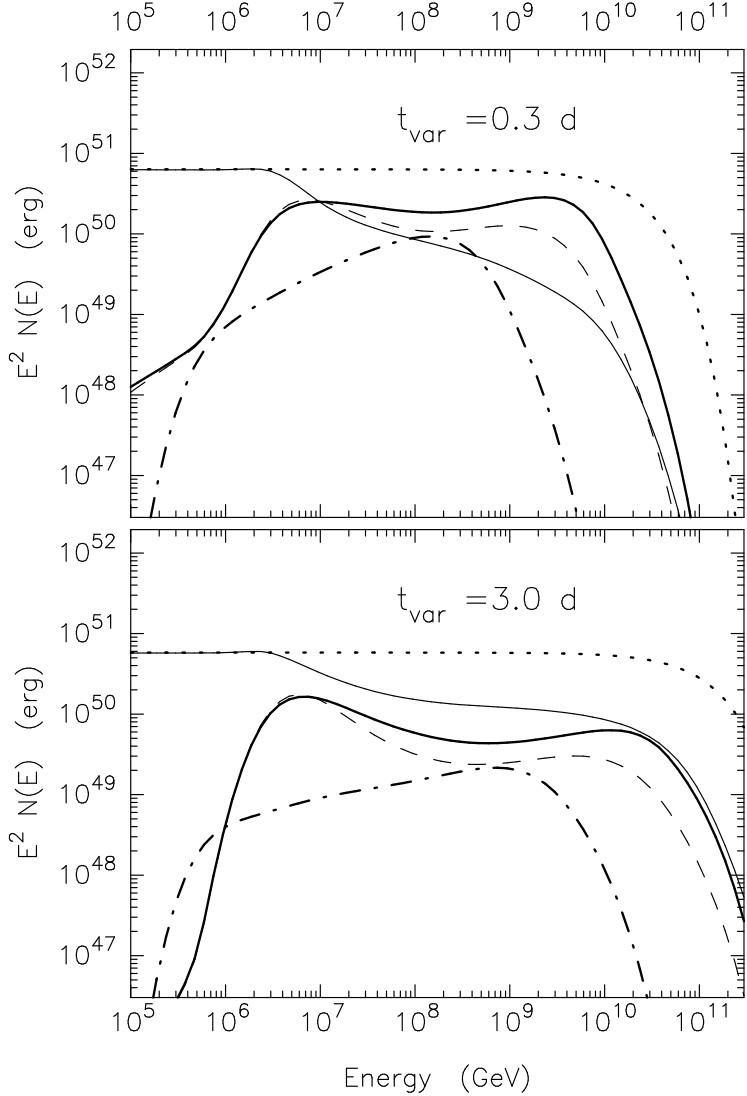}
\caption{Spectra of injected protons (dotted curves), protons surviving
 when the blob reaches the edge of the BLR (thin solid curves),
neutrons that escape from the blob (thick solid curves), neutrons
(dashed curves) and gamma-rays (dot-dashed curves) that reach the edge
of the BLR without attenuation for the same overall parameters as used
in Figure 1 for $\delta = 10$, but calculated assuming different sizes
of the blob, corresponding to $t_{var} = 0.3$ and 3 days in the upper
and lower panels, respectively. In calculation here we assume also 
that the same amount of accelerated  protons as in Figure 1 is now injected 
continuously during the entire time of passage of the jet through BLR,
$t_{inj}=7.9\,\rm d$ in the observer frame, instead of $t_{inj}=2\,\rm d$
in Figure 1. }
\label{f4}
\end{figure}

\clearpage

\begin{figure}
\plotone{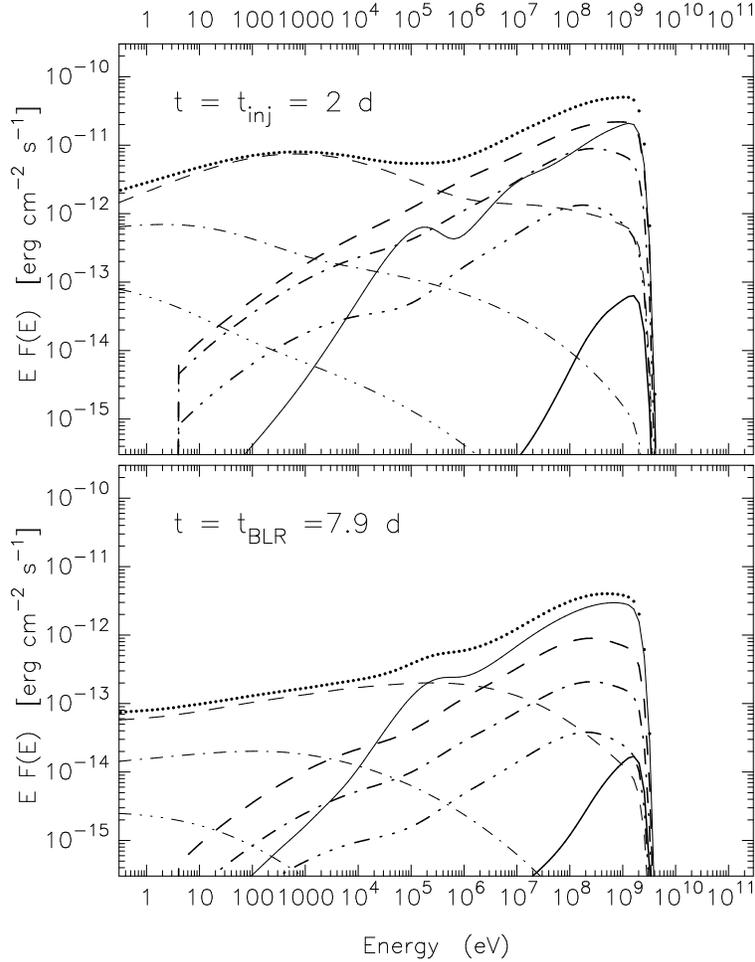}
\caption{Radiation flux produced in and escaping from the blob (full dots) 
following the electromagnetic cascade initiated by energetic electrons
and gamma rays produced in photopion interactions for the case $\delta
= 10$ and the model parameters as in Figure 1.  
The thick and thin curves correspond to synchrotron and Compton
scattered radiation, respectively. The radiation of the first
generation of electrons, which includes both the electrons from
$\pi^\pm$ decay and the electrons produced by absorption of
$\pi^0$-decay gamma rays in the blob, are shown by the solid curves.
The dashed, dot-dashed and 3-dot--dashed curves show contributions from
the 2d, 3d and 4th generations of cascade electrons, respectively.
The upper and lower panels (a) and (b) show the cascade radiation flux
at t = 2d, when injection of accelerated protons 
stops, and at $t=7.9\,\rm d$, when the blob reaches the edge of
 the BLR, respectively.}
\label{f5}
\end{figure}

\clearpage

\begin{figure}
\epsscale{1.0}
\plotone{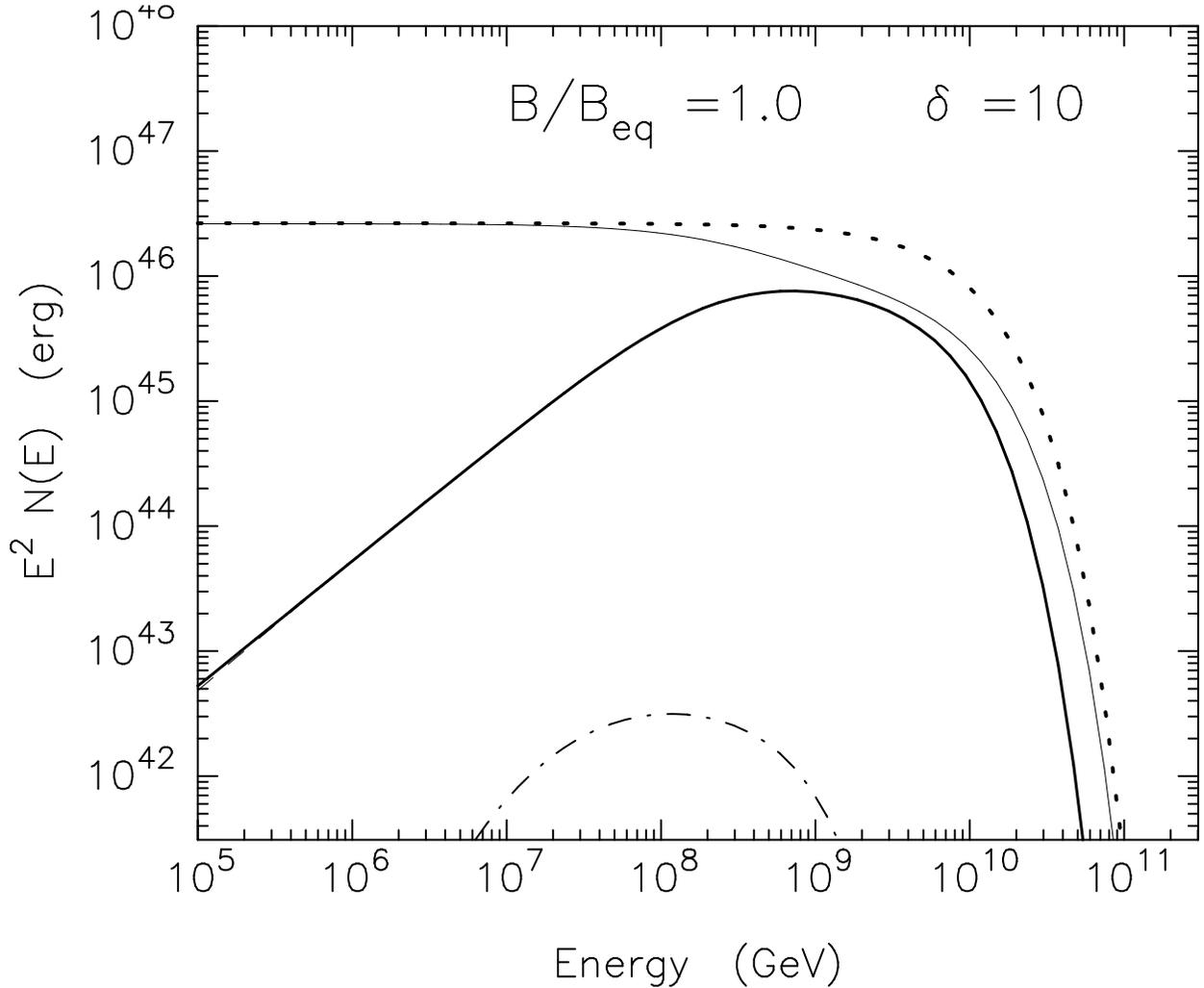}
\caption{Total energy spectra of injected protons (dotted curve), and protons 
remaining in the blob (thin solid curve)  at the time when the jet, with model 
parameters relevant to
the April 1997 flare from Mrk 501 (see Section 3.2),  $t_{var} = 1\,$d and 
$\delta = 10$, is at a distance $R = 0.4$ pc from the center. The thick solid
curve shows the total spectrum of the escaping neutrons, and the 
dot-dashed curve shows the spectrum of gamma rays
that could be produced by these neutrons outside the
 blob and escape the scattering region.  }
\label{f6}
\end{figure}

\clearpage

\begin{figure}
\epsscale{0.6}
\plotone{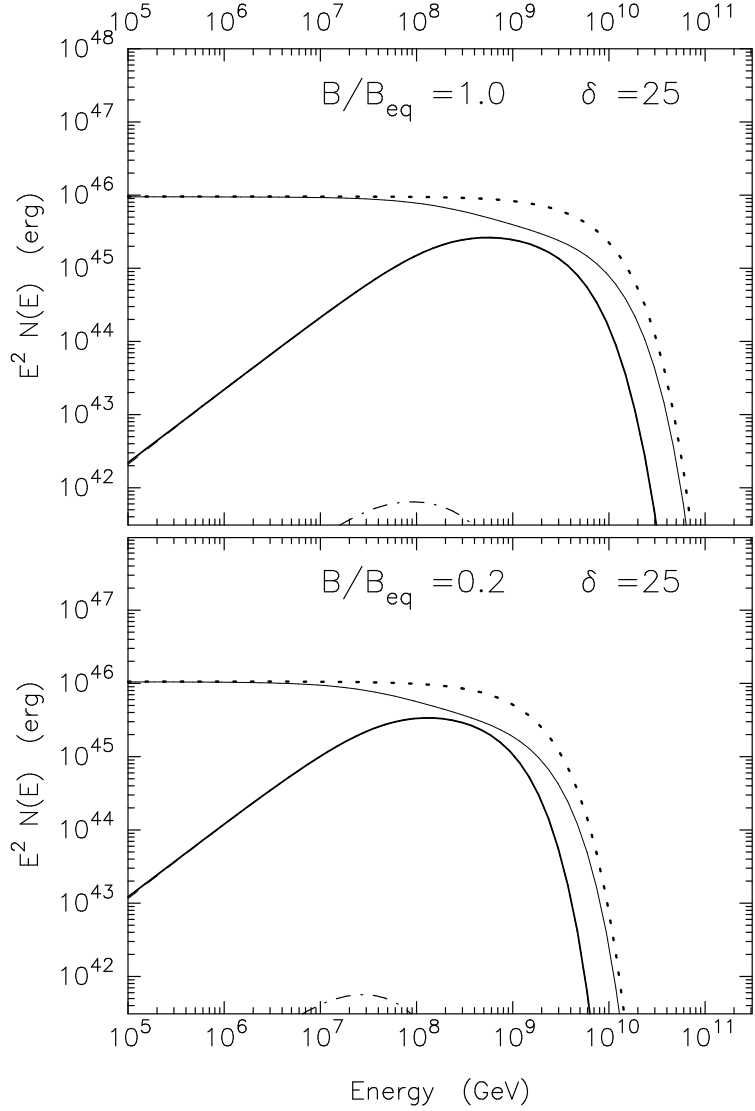}
\caption{Energy spectra of protons neutrons, and gamma rays from a jet
in Mrk\,501 type object, as in Figure 
6, but calculated for the case $t_{var} = 0.1\,$d and 
$\delta = 25$, with $B = B_{eq} = 1.0$  in (a), and $B = 0.2 B_{eq}$ in (b).}
\label{f7}
\end{figure}

\clearpage

\begin{figure}
\epsscale{1.0}
\plotone{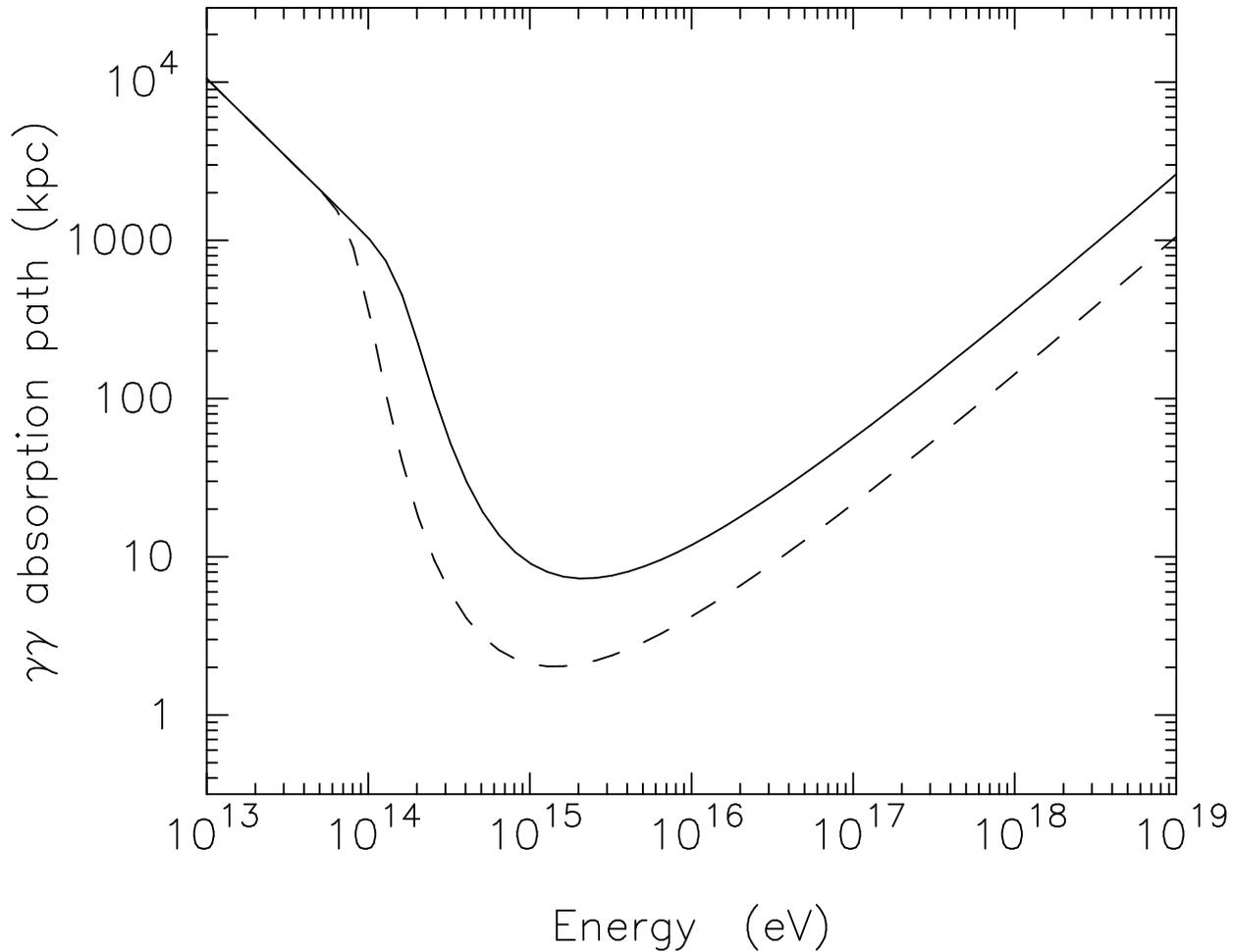}
\caption{Mean-free-path of gamma rays due to pair production interactions with 
photons of the diffuse cosmic microwave background radiation for $z =
0$ (solid curve) and $z = 0.538$ (dashed curve), including the diffuse
extragalactic infrared radiation field (see Section 4.3).}
\label{f8}
\end{figure}

\end{document}